\documentclass[onecolumn,11pt,a4paper,aps]{revtex4}

\usepackage{colordvi}
\usepackage{graphicx,sidecap}
\usepackage{bm}
\usepackage{braket}
\usepackage{amssymb}
\usepackage{mathrsfs}
\usepackage{amsmath}
\usepackage{xcolor}
\usepackage{hyperref}
\usepackage{enumerate}
\usepackage[toc,title]{appendix}
\usepackage{verbatim}
\hypersetup{colorlinks,bookmarksopen,bookmarksnumbered,
 citecolor=red,
 linkcolor=blue,
 pdfstartview=green,
 urlcolor=purple}

\newcommand{\eq}[1]{\begin{equation}#1\end{equation}}
\newcommand{\dd}{\mathrm{d}}
\newcommand{\ex}{\mathrm{e}}

\def\be{\begin{equation}}
\def\ee{\end{equation}}
\def\bea{\begin{eqnarray}}
\def\eea{\end{eqnarray}}

\newcommand{\betax}{\beta(x)}
\newcommand{\betabix}{\tilde \beta(x)}

\begin{document}

\title {Local and non-local properties of the entanglement Hamiltonian \\ for two disjoint intervals}
\author{ Viktor Eisler$^1$, Erik Tonni$^2$ and Ingo Peschel$^3$}

\affiliation{
$^1$Institut f\"ur Theoretische Physik, Technische Universit\"at Graz, Petersgasse 16,
A-8010 Graz, Austria\\
$^2$SISSA and INFN Sezione di Trieste, via Bonomea 265, I-34136 Trieste, Italy\\
$^3$Fachbereich Physik, Freie Universit\"at Berlin, Arnimallee 14, D-14195 Berlin, Germany
}

\begin{abstract}
  We consider free-fermion chains in the ground state and the entanglement Hamiltonian for a
  subsystem consisting of two separated intervals. In this case, one has a peculiar long-range
  hopping between the intervals in addition to the well-known and dominant short-range hopping.
  We show how the continuum expressions can be recovered from the lattice results for
  general filling and arbitrary intervals. We also discuss the closely related case of a
  single interval located at a certain distance from the end of a semi-infinite chain and the
  continuum limit for this problem. Finally, we show that for the double interval in the continuum
  a commuting operator exists which can be used to find the eigenstates.

\end{abstract}
\maketitle
%%%%%%%%%%%%%%%%%%%%%%%%%%%%%%%%%%%%%%%%%%%%%%%%%%%%%%%%%%%%%%%%%%%%%%%%%%%%%%%%%%%%%%%%%%%%%
\section{Introduction}
\label{sec:intro}

In entanglement studies, one divides a quantum system into two parts and determines how
they are coupled in the chosen state. This information is encoded in the reduced 
density matrix $\rho$ of one of the subsystems or, writing $\rho=\exp(-\mathcal{H})/Z$, in
the operator $\mathcal{H}$. The latter has therefore been called the entanglement Hamiltonian,
while in quantum field theory the name modular Hamiltonian is often used
\cite{Bisognano/Wichmann75, Bisognano/Wichmann76, Hislop/Longo82, Haag-book}. 
Its form depends on the quantum state in question as well as on the type of partition but also shows certain
universal features and it has been the topic of various studies in recent years. For a
review, see \cite{Dalmonte/Eisler/Falconi/Vermersch22}.

In one dimension, which we consider here, there are two standard partitions:  a division of
a chain (or line) into two halves or into a segment (or interval) and the remainder. In both cases,
there is a simple analytical result for $\mathcal{H}$ if the chain is infinite and one is
dealing with a continuous critical system in its ground state. Then 
\begin{equation}
 \mathcal{H}\,=\,2\pi \int_A \textrm{d}x\,\beta(x)\,T_{00}(x)\,,
\label{general}
\end{equation}
where the integration runs over the subsystem $A$, $T_{00}(x)$ is the energy density in
the physical Hamiltonian and $\beta(x)$ is a weight function. For a half-infinite subsystem
occupying the region $x > 0$, it varies linearly, $\beta(x)=x$, while for critical systems and an interval between
$x=a$ and $x=b$, it is given by the parabola $\beta(x)=(x-a)(b-x)/(b-a)$. 
The first result is due to Bisognano and Wichmann \cite{Bisognano/Wichmann75,Bisognano/Wichmann76}
and the second one can be obtained from it by a conformal mapping, 
see \cite{Hislop/Longo82, Casini/Huerta/Myers11, Wong_etal13, Cardy/Tonni16}. 
This expression has two characteristic features: $\mathcal{H}$ contains only local terms as
the physical Hamiltonian and it is inhomogeneous. More precisely, the terms are large in the
interior of the subsystem and small near its boundary where the factor $\beta(x)$ vanishes
linearly.

For discrete systems, the situation is somewhat more complicated. For free fermions 
on a lattice, one finds for the interval a dominant nearest-neighbour hopping in $\mathcal{H}$
which does not quite vary parabolically, but also hopping to more distant neighbours with
smaller amplitudes \cite{Peschel/Eisler09,Eisler/Peschel17}. Thus the entanglement Hamiltonian
is not strictly local. However, it has been shown numerically \cite{Arias_etal17_1} and also
analytically \cite{Eisler/Tonni/Peschel19} that in the continuum limit one recovers the conformal
result for $\beta(x)$ by properly including the longer-range terms. 
For free massless bosons in the form of coupled harmonic oscillators,
the same was found through a numerical approach \cite{DiGiulio/Tonni19},
and a similar analysis was carried out in higher dimensions for a spherical domain \cite{Javerzat/Tonni21}.

An intriguing new feature appears if the subsystem consists of two (or more) disjoint intervals. This
was discovered by Casini and Huerta in a study of massless Dirac fermions, i.e. free fermions in the
continuum \cite{Casini/Huerta09} and later investigated further in
\cite{Longo/Martinetti/Rehren09, Rehren/Tedeco13, Arias_etal17_1,Arias_etal18,Wong19}.
One then finds a peculiar long-range coupling between each point in one interval and one
single partner in the other interval. Mathematically, it results from the form of the
eigenfunctions of the correlation kernel from which one can construct the entanglement
Hamiltonian. They show spatial oscillations $\mathrm{exp}( \textrm{i} pw(x))$ in a variable $w(x)$ which
depends logarithmically on $x$, and as a result a function $\delta'(w(x)-w(y))$ appears in the
expression for $\mathcal{H}$ which gives contributions not only for $y=x$ but also for a point
$y=x_{\textrm{\tiny c}}$ located in the other interval. The locus of the coupled points is a
hyperbola in the $(x,y)$ plane while the amplitude of the coupling varies with $x$ and is
given by a function  $\tilde\beta(x)$ resembling the $\beta(x)$ for the local terms. 

The situation on the lattice is again more complicated. To some extent, it was studied in
\cite{Arias_etal17_1} for slightly off-critical systems. Here we will investigate it in 
more detail and strictly for the critical case. In $\mathcal{H}$ one then finds hopping between
a site and many others in the other interval. In the Hamiltonian matrix, this corresponds to
characteristic regions of elements which are larger than those in the surrounding and do
not scale with the size of the intervals. The shape of the regions is reminiscent of the
hyperbolae in the continuum. We show that, in analogy with the short-range hopping, a continuum
limit can be taken in which they combine and give the continuum result for $\tilde\beta(x)$.
This is done first for a half-filled system and symmetric intervals and then generalized
to other fillings and unequal intervals.

We also treat a geometry which corresponds to some extent to one-half of the symmetric
double interval, namely a single interval located at some distance from the boundary of a
half-infinite chain. Its continuum version for Dirac fermions was studied recently in
\cite{Mintchev/Tonni21}, and it turned out that a long-range coupling also exists there,
but now inside the single interval. On the lattice, this is a certain drawback since now
the usual longer-range terms and the new ones appear in the same region of the entanglement Hamiltonian
matrix. Nevertheless, we were able to separate them by a proper choice of the summations
in the continuum limit. Thus we could reobtain the continuum weight factor $\tilde\beta(x)$
also in this case, although only for half filling.

Finally, we consider an aspect which plays a considerable role in the treatment of single
intervals in free-fermion chains. Namely, a simple operator exists in a number of cases
which commutes with the correlation matrix (or kernel) and thus has the same eigenfunctions,
see \cite{Eisler/Peschel13,Eisler/Peschel18,CNV19,CNV20}. For the hopping model, this operator
first led to the logarithmic oscillations of these functions \cite{Peschel04} and later
to analytical expressions for
the matrix elements in $\mathcal{H}$ \cite{Eisler/Peschel17}. Here we show that such an
operator also exists in the continuum case for both a single and a double interval.
In the first case, it is a simple first-order differential operator, and in the second case it
contains an additional difference term. A similar operator can also be found for an interval on the half line. 

The layout of the paper is as follows. In section 2 we describe the setting and the models
and give some known results for the double interval. In section 3 we consider the hopping
model and present numerical results for the matrix elements in $\mathcal{H}$ for rather
large intervals, obtained as usual from high-precision diagonalizations of the reduced
correlation matrix. In section 4 we describe the continuum limit for the non-local terms in
$\mathcal{H}$ and compare with the continuum result. In section 5, after presenting some
numerical results, we do the same for a single interval in a half-infinite hopping chain.
In section 6 we present the commuting operator. Finally, in section 7, we sum up our
findings and give some outlook.
In two appendices we construct the eigenfunctions of the commuting differential operator
and derive its form for the interval on the half line.
%correlation kernel from it
%treat the case of large separations of the intervals and collect some useful expressions. 

%%%%%%%%%%%%%%%%%%%%%%%%%%%%%%%%%%%%%%%%%%%%%%%%%%%%

\section{Setting}
  \label{sec:setting}

We describe here the two situations for which we consider the entanglement Hamiltonian.

\subsection{Continuum results\label{sec:cont}}

The case of one-dimensional Dirac fermions with zero mass has been treated repeatedly in
the past. One then is dealing with right- and left-moving particles described by field
operators $\psi_{\textrm{\tiny R}}(x)$ and $\psi_{\textrm{\tiny L}}(x)$ and the Hamiltonian
can be written
\be
\hat H= \int \textrm{d} x \,T_{00}(x)\,,
\label{H_cont-def}
\ee
with the energy density $T_{00}(x)$ given by
\be
\label{T00-def}
T_{00}(x) 
\,\equiv\,
\,\frac{\textrm{i}}{2}
:\! \!
\Big[ \Big (
\psi^\dagger_{\textrm{\tiny R}}\, (\partial_x \psi_{\textrm{\tiny R}}) 
-
(\partial_x \psi^\dagger_{\textrm{\tiny R}})\, \psi_{\textrm{\tiny R}}
\Big )(x)
- 
\Big(
\psi^\dagger_{\textrm{\tiny L}}\, (\partial_x \psi_{\textrm{\tiny L}})
-
(\partial_x \psi^\dagger_{\textrm{\tiny L}})\,  \psi_{\textrm{\tiny L}}
\Big) (x)
\Big]\!\! : \;.
\ee
In the ground state, all levels with negative energy are occupied and the correlation
function for the right-movers is
\be
C_{\textrm{\tiny R}}(x,y)
=
\langle \psi^\dag_{\textrm{\tiny R}}(x)\psi_{\textrm{\tiny R}}(y) \rangle 
 =
 \int_{-\infty}^{0} \frac{\textrm{d} q}{2\pi}\,\mathrm{exp}(-\textrm{i}q(x-y)+q\epsilon)
 =
 \frac{\pi}{2}\delta(x-y)+\frac{\textrm{i}}{x-y}\,,
\label{corrfct-def}
\ee
and similarly for the left-movers.

The entanglement Hamiltonian can then be obtained from the eigenfunctions of these
integral kernels \cite{Mush-book}
as sketched below for the discrete case. For a double interval
$A = A_1 \cup A_2$ consisting of two disjoint intervals  $A_1=(a_1,b_1)$ and $A_2=(a_2,b_2)$,
it was found in \cite{Casini/Huerta09} that
\be
\label{H_A-def-sum-loc-bi-loc}
\mathcal{H} \,=\, \mathcal{H}_{\textrm{\tiny loc}}  + \mathcal{H}_{\textrm{\tiny bi-loc}} \,,
\ee
with the local term as in (\ref{general}) and the bi-local term given by
%\be
%\label{K_A-local-def}
%\mathcal{H}_{\textrm{\tiny loc}} 
%\,=\,
%2\pi 
%\int_A 
%\beta_{\textrm{\tiny loc}}(x) \, T_{00}(x)\, \textrm{d} x
%\ee
\be
\label{H_A-bi-local-def}
\mathcal{H}_{\textrm{\tiny bi-loc}} 
\,=\,
2\pi 
\int_A  \textrm{d} x \,
\tilde\beta(x) \, T_{\textrm{\tiny bi-loc}}(x, x_{\textrm{\tiny c}} ) \,,
\ee
where $T_{\textrm{\tiny bi-loc}}$ is the following operator 
\be
\label{T-bilocal-2int}
T_{\textrm{\tiny bi-loc}}(x, y) 
 \,\equiv \,
\frac{\textrm{i}}{2}\;
\bigg\{ 
\!:\!\!\Big[\, \psi^\dagger_{\textrm{\tiny R}}(x) \,  \psi_{\textrm{\tiny R}}(y) - \psi^\dagger_{\textrm{\tiny R}}(y) \,  \psi_{\textrm{\tiny R}}(x) \, \Big]\!\!: 
-
:\!\! \Big[\,  \psi^\dagger_{\textrm{\tiny L}}(x) \,  \psi_{\textrm{\tiny L}}(y) - \psi^\dagger_{\textrm{\tiny L}}(y) \,  \psi_{\textrm{\tiny L}}(x) \, \Big] \!\!: \!
\bigg\} \;.
\ee
The conjugate point $x_{\textrm{\tiny c}}$ in (\ref{H_A-bi-local-def}) is defined in terms of $x$ as
\be
\label{x-conjugate-2-int}
x_{\textrm{\tiny c}} 
\,\equiv\,
\frac{(b_1 b_2 - a_1 a_2)\, x + (b_1 + b_2) a_1 a_2 - (a_1 + a_2) b_1 b_2 }{(b_1 + b_2 - a_1 - a_2)\, x + a_1 a_2 - b_1 b_2}
\,=\,
x_0 - \frac{R^2}{x - x_0} \,,
\ee
where
\be
x_0 \equiv \frac{b_1 b_2 - a_1 a_2}{b_1 - a_1 + b_2 - a_2}\,,
\;\;\;\;\qquad\;\;\;\;
R^2 \equiv
\, \frac{(b_1 - a_1)\,(b_2 - a_2)\, (b_2 - a_1)\, (a_2 - b_1)}{(b_1 - a_1 + b_2 - a_2)^2}\,,
\ee
and lies in $A_2$ if $x$ is in $A_1$ and vice versa. It has
a simple geometrical meaning, namely $x_{\textrm{\tiny c}}$ is the reflection of $x$ on a circle with
radius $R$ around $x_0$ plus a reflection with respect to $x_0\in \left(b_1,a_2\right)$.

In order to write down the weight functions $\beta(x)$ and $\tilde\beta(x)$
occurring  in the local term (\ref{general}) and in the bi-local term (\ref{H_A-bi-local-def}),
respectively, it is convenient to use the function
\be
\label{w-function-def}
w(x) \,=\, \ln\!\left[ - \frac{(x - a_1)\,(x - a_2)}{(x - b_1) \, (x - b_2)} \right] \,,
\ee
which occurs in the solution of the eigenvalue problem and has the property
$w(x_{\textrm{\tiny c}})=w(x)$, which actually defines $x_{\textrm{\tiny c}}$.
In terms of (\ref{w-function-def}) one has 
\be
\label{beta-loc-w-def}
\beta(x) =  \frac{1}{w'(x)} \,,
\;\;\;\;\qquad\;\;\;\;
\tilde\beta(x)
\, = \,
\frac{\beta(x_{\textrm{\tiny c}} )}{ x - x_{\textrm{\tiny c}} } \,.
\ee
%and
%\be
%\label{beta-bi-loc-w-def}
%\beta_{\textrm{\tiny bi-loc}}(x)
%\, = \, 
%\frac{\beta(x_{\textrm{\tiny c}} )}{ x - x_{\textrm{\tiny c}} }
%\ee

In the special case of the symmetric configuration $A_1= (-b, -a), \,A_2= (a,b)$, where $b > a > 0$,
all formulae simplify. Then $x_0=0$ and $R^2=ab$, the conjugate point becomes $x_{\textrm{\tiny c}} = - \tilde x$
where $\tilde x= ab/x$ and the functions in (\ref{beta-loc-w-def}) are
\be
\label{beta-loc-2int-sym}
\beta(x)
= 
\frac{(b^2 - x^2)\,(x^2 - a^2)}{2\,(b-a)\, (a\,b +x^2)} \,,
\ee
and
\be
\label{beta-bi-loc-2int-sym}
\tilde\beta(x)
= \frac{a\,b}{x\,(a\,b+x^2)}\,\, \beta(x) \,.
%\frac{a\, b \, (b^2-x^2) \, (x^2-a^2)}{2\,(b-a)\, x\,(a\, b + x^2)^2} 
\ee

Note that $\beta(x)$ is related to the corresponding quantities for the single
intervals $A_1$ and $A_2$ via $1/\beta=1/\beta_1+1/\beta_2$. In the following, we will compare
the lattice results to these expressions and in the course of this also show graphs of them.

\subsection{Lattice model}

On the lattice, we will study an infinite fermionic hopping chain with Hamiltonian
\eq{
\hat H = 
- \sum_n t \, (c^{\dag}_n c_{n+1} + c^{\dag}_{n+1} c_{n})
+\mu \sum_n c^{\dag}_n c_n \, ,
\label{Hff}
}
with fermionic creation/annihilation operators $c_n^\dag$ and $c_n$ and chemical potential $\mu$.
Setting $t=1/2$ and $\mu=\cos q_F$, the ground state is a Fermi sea with occupied momenta
$q\in\left[-q_F,q_F\right]$. Similarly to the continuum case, we consider two
disjoint segments $A_1$ and $A_2$ containing $N_1$ and $N_2$ lattice sites, respectively,  
and separated by a distance of $D$ sites. The entanglement Hamiltonian for the combined
subsystem $A = A_1 \cup A_2$ is then a quadratic expression in the fermionic operators
\cite{Peschel03,Peschel/Eisler09}
\eq{
\mathcal{H}=  \sum_{i,j \in A} \, H_{i,j} \,c^{\dag}_i c_j \,,
\label{EHff}
}
where the matrix $H$ is given by
\eq{
H_{i,j}= \sum_{k=1}^{N_1+N_2}\,\phi_k(i)\; \varepsilon_k\; \phi_k(j) \, , 
\;\;\;\;\qquad \;\;\;\;
\varepsilon_k = \ln \frac {1-\zeta_k}{\zeta_k} \, ,
\label{Hij}}
via the eigenvalues $\zeta_k$ and eigenvectors $\phi_k(i)$ of the reduced correlation
matrix $C_A$. This is composed of matrix elements $C_{i,j}=\langle c_i^\dag c_j \rangle$
given explicitly by
\be
C_{i,j}=\frac{\sin[q_F(i-j)]}{\pi(i-j)} \,,
\label{corr-lattice}
\ee
with indices restricted to $i,j\in A$. Due to the two subintervals, $C_A$ and $H$ have
a block structure, which for the latter can be represented as
\eq{
H = \left(
\begin{array}{c|c}
H^{\textrm{\tiny (1)}}  & H^{\textrm{\tiny (1,2)}}
\\ \hline
H^{\textrm{\tiny (2,1)}}  & H^{\textrm{\tiny (2)}}
\end{array}
\right).
\label{Hblock}
}
The diagonal blocks $H^{\textrm{\tiny (1)}} $ and $H^{\textrm{\tiny (2)}} $ describe
hopping {\it{within}} the corresponding segment, whereas the off-diagonal ones contain
long-range hopping terms {\it{between}} the two segments and satisfy
$H^{\textrm{\tiny (2,1)}}_{i,j}=H^{\textrm{\tiny (1,2)}}_{j,i}$.

In the following, we present
the results of numerical calculations of $H$ for rather large subsystems with values of
$N_1$ and $N_2$ up to 160.
As is well known, this requires an extreme accuracy in the
diagonalization procedure and amounts to working with up to several hundreds of decimal places in
{\it{Mathematica}}.

%%%%%%%%%%%%%%%%%%%%%%%%%%%%%%%%%%%%%%%%%%%%%%%%%%%%%%%%%%%%%%%%%%%%%%%%%%%%%%%%%%%%%%%%%
\section{Entanglement Hamiltonian on the lattice}
\label{sec:lattice}
%%%%%%%%%%%%%%%%%%%%%%%%%%%%%%%%%%%%%%%%%%%%%%%%%%%%%%%%%%%%%%%%%%%%%%%%%%%%%%%%%%%%%%%%
% comment on numerics

\begin{figure}[t]
\centering
\vspace{-1cm}
\includegraphics[width=0.45\textwidth]{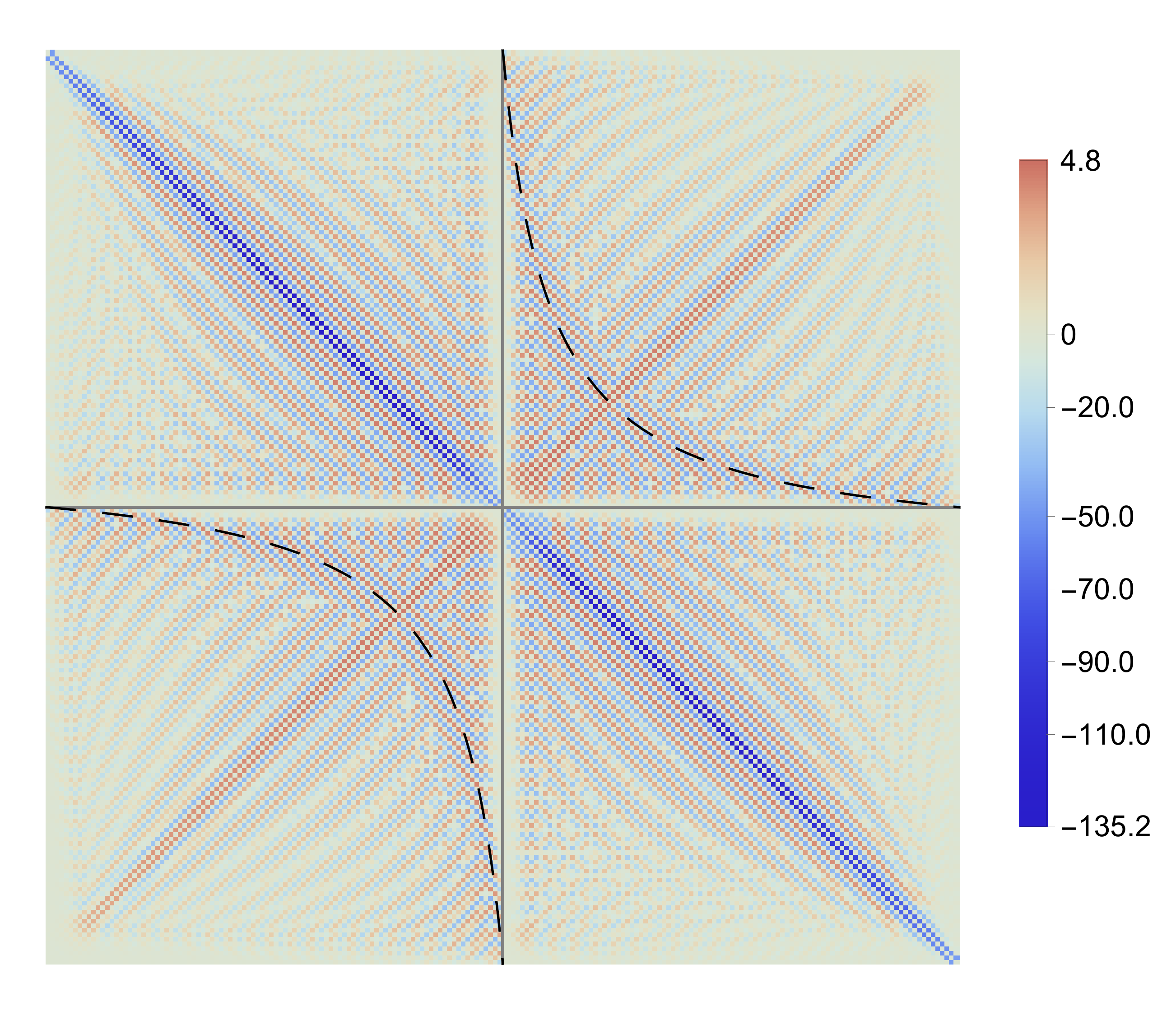}
\qquad
\raisebox{0.5cm}{\includegraphics[width=0.49\textwidth]{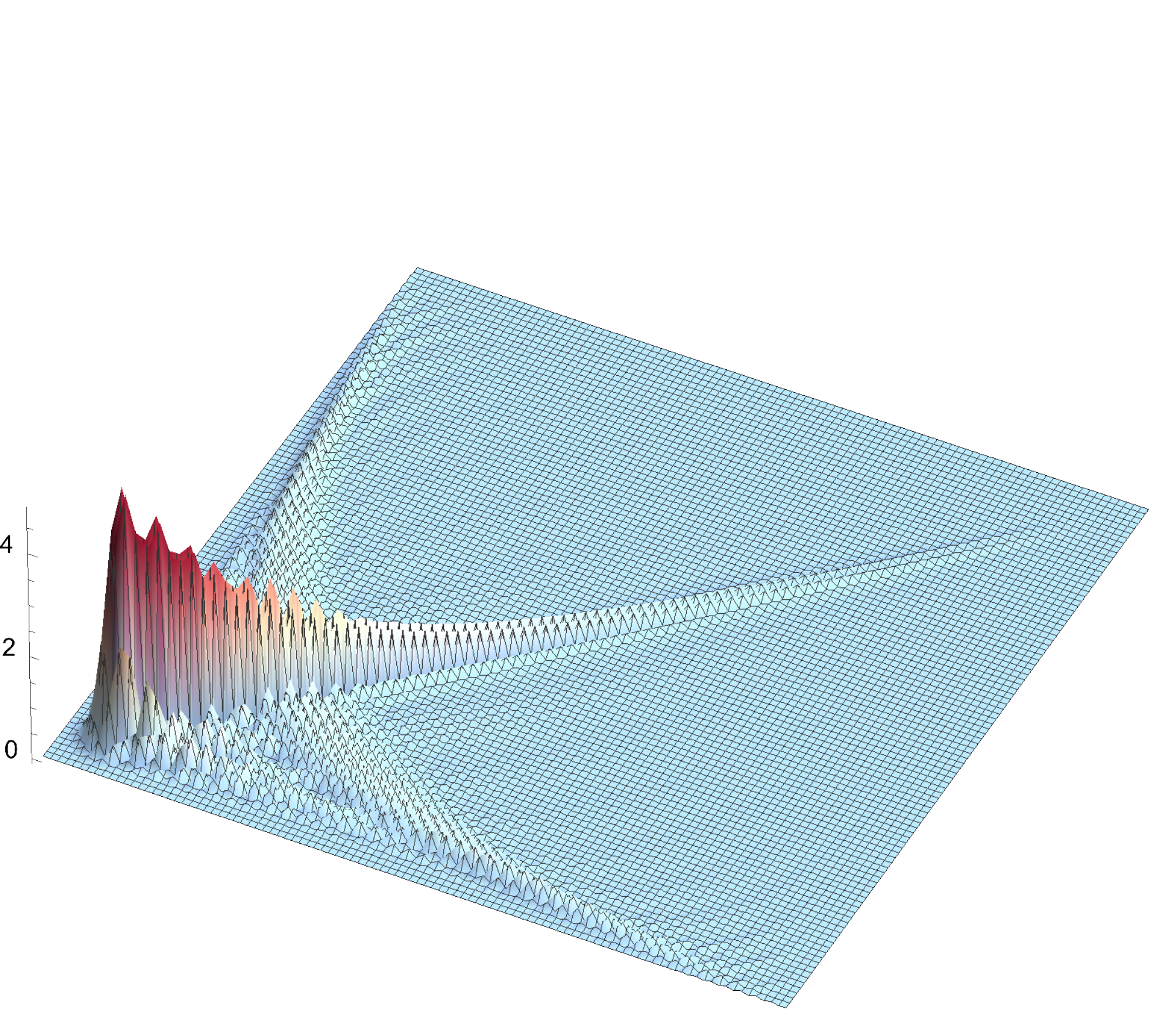}}

\vspace{-.8cm}
\includegraphics[width=0.45\textwidth]{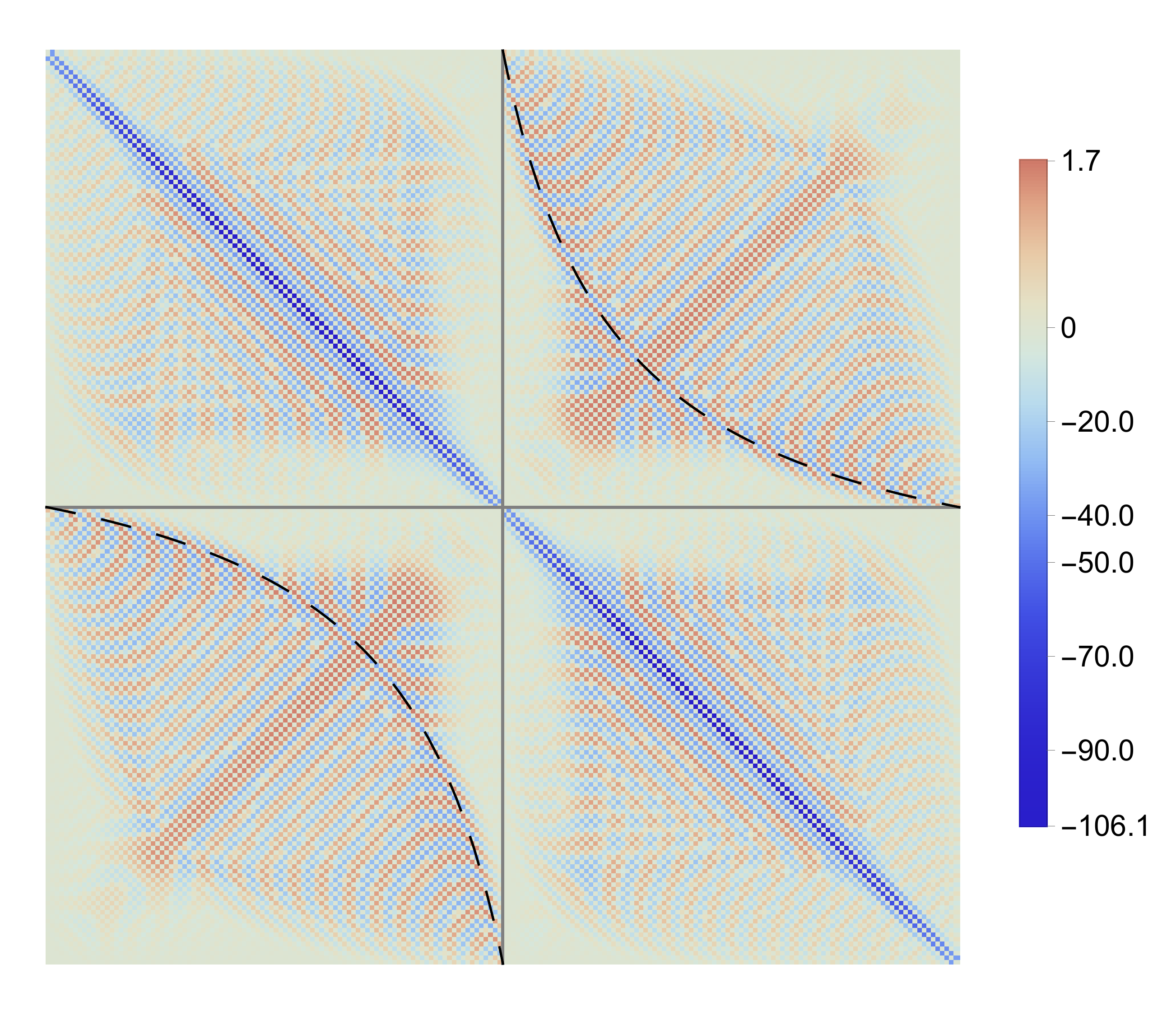}
\qquad
\raisebox{0.5cm}{\includegraphics[width=0.49\textwidth]{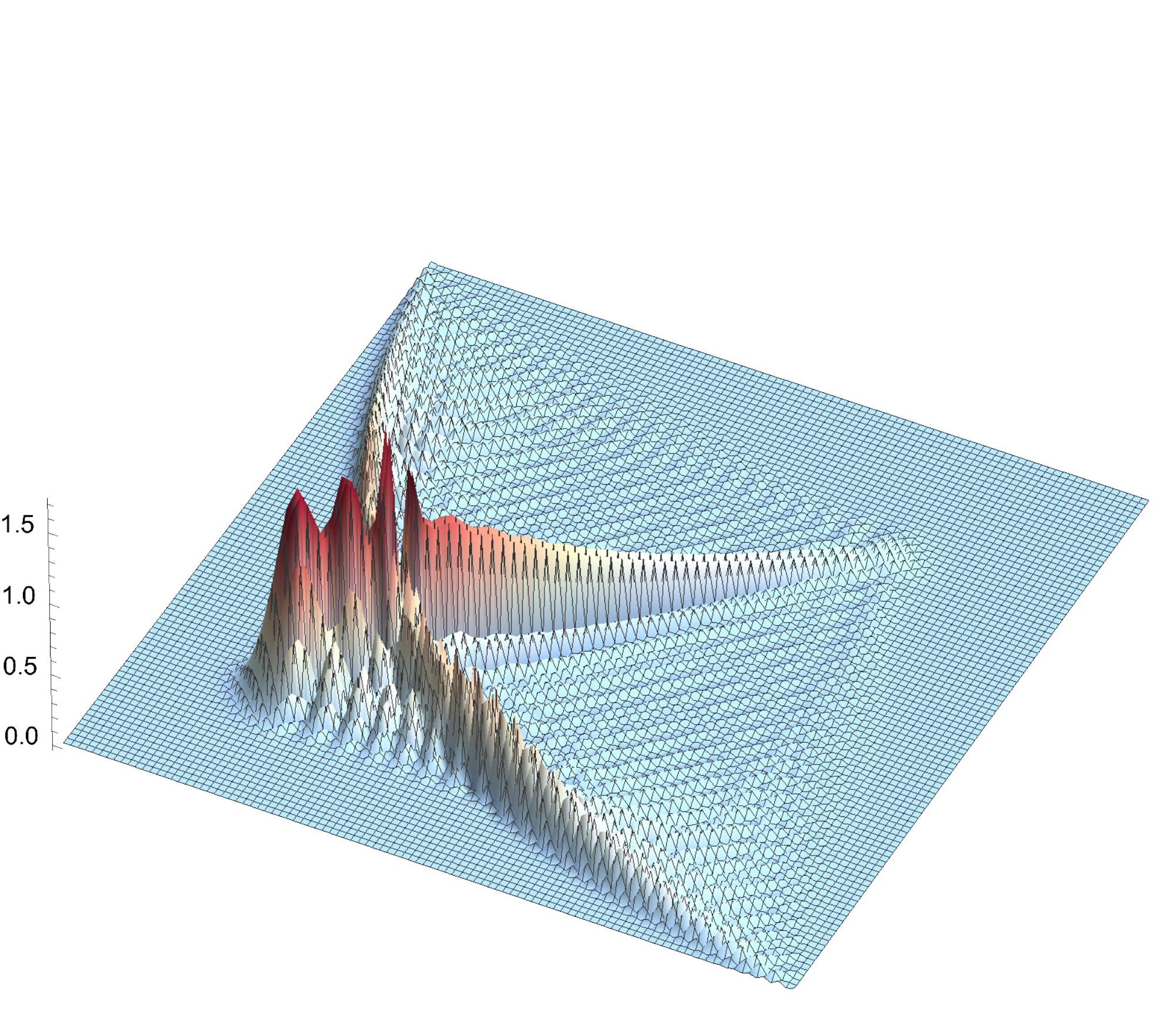}}

\vspace{-.8cm}
\includegraphics[width=0.45\textwidth]{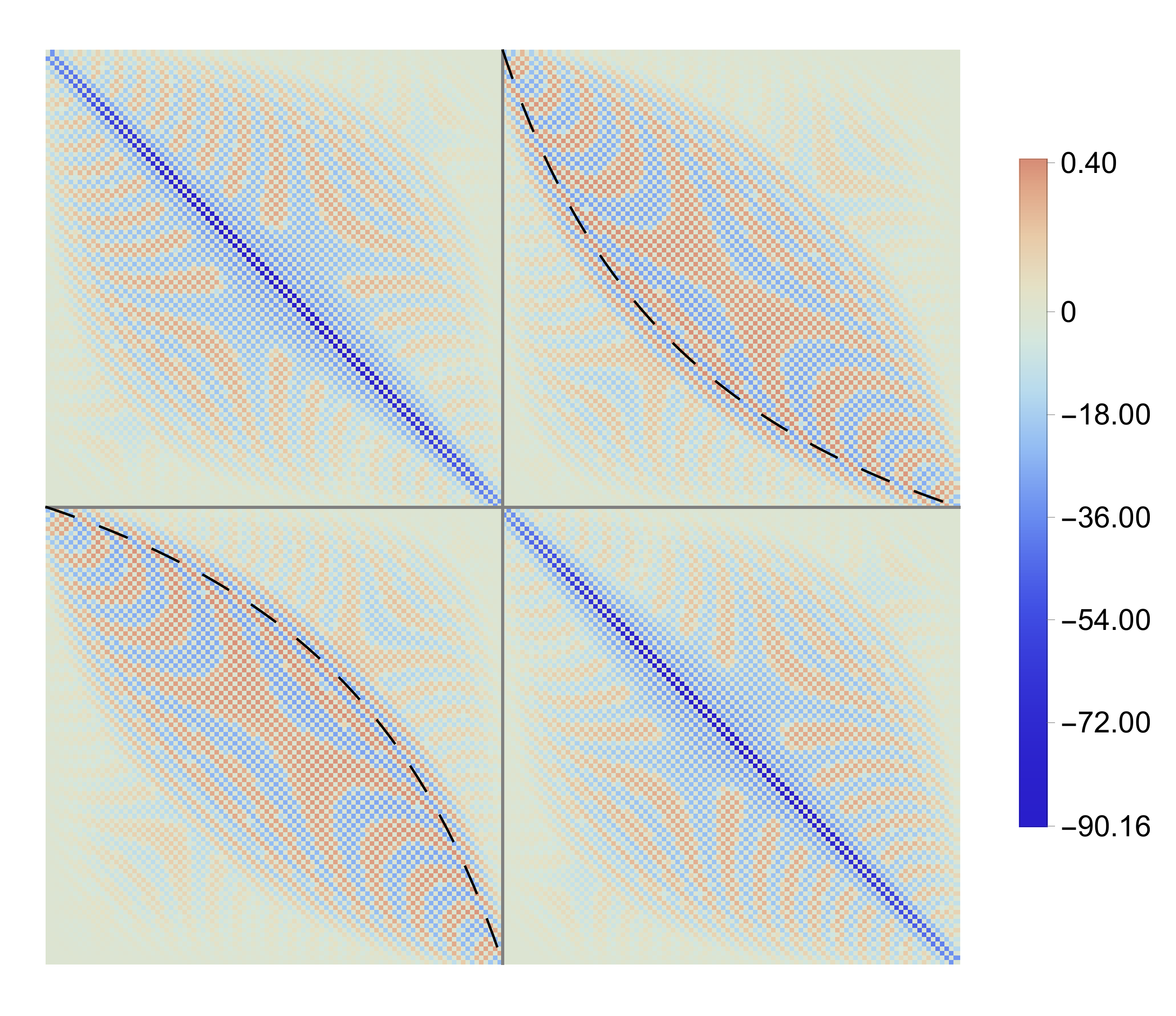}
\qquad
\raisebox{0.5cm}{\includegraphics[width=0.49\textwidth]{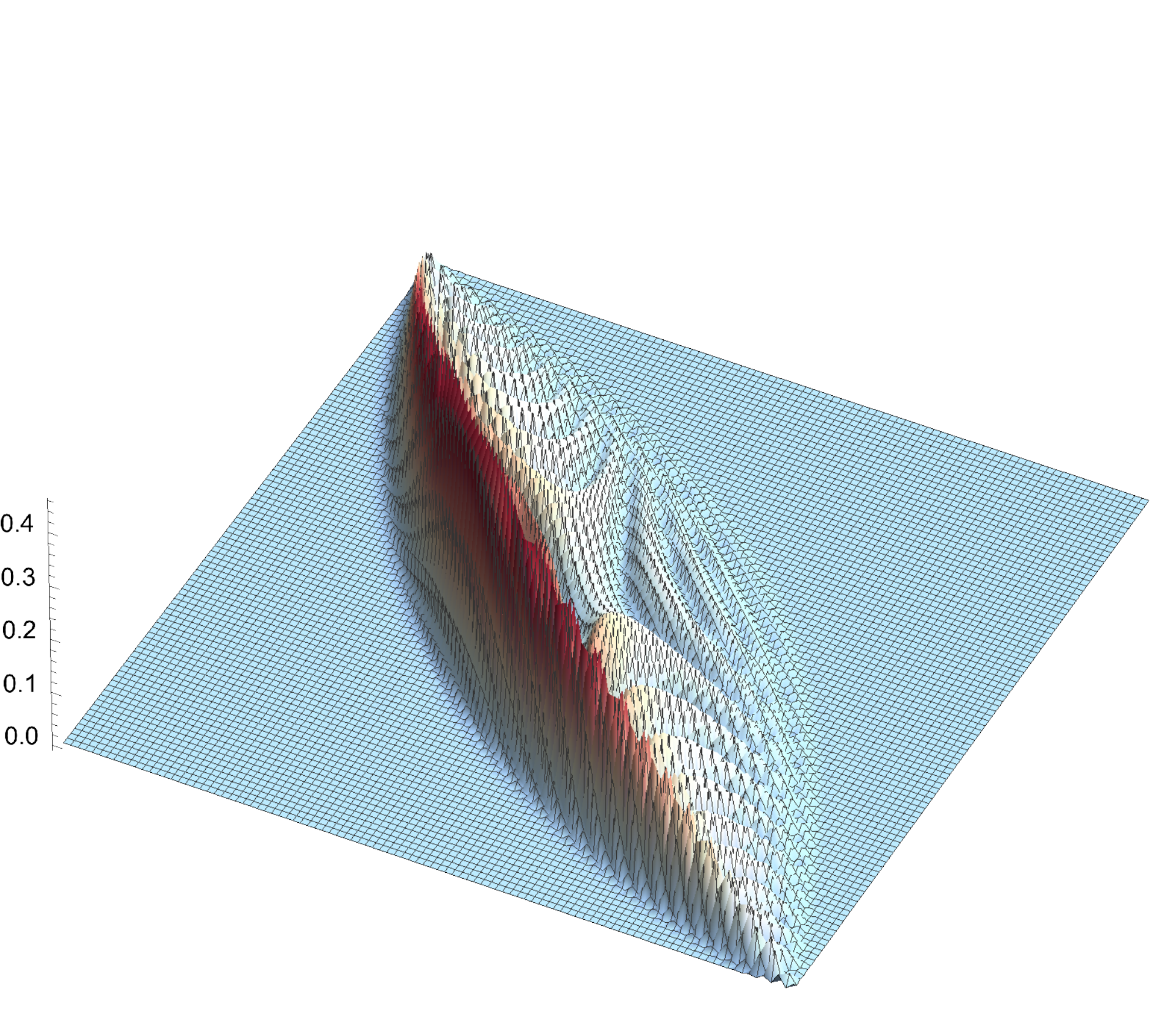}}

\vspace{0cm}
\caption{Left: Visualization of the matrix elements in $H$ for equal intervals of size $N=100$ and separations
$21,51,101$, from top to bottom, with the color code representing the amplitudes.
The four blocks in \eqref{Hblock} are separated by grey lines, and the black dashed lines in the
off-diagonal blocks indicate the hyperbolae $x_{\textrm{\tiny c}}(x)$ in \eqref{x-conjugate-2-int}.
Right: 3D plots of the off-diagonal block $H^{\textrm{\tiny (1,2)}}$,
with the absolute values of the matrix elements shown. The color coding differs from that on the left.}
\label{fig:EHmatrix}
\end{figure}

To get a first impression of the detailed structure of $H$, we visualize its matrix elements
using color coded plots in Fig.\,\ref{fig:EHmatrix}. For simplicity, we consider a half-filled chain
and equal intervals of size $N_1=N_2=100$ separated by a distance $21,51$ and $101$ respectively.
On the left hand side of Fig.\,\ref{fig:EHmatrix}, a 2D-plot of the full matrix $H$ is shown,
where each dot corresponds to a matrix element and its magnitude is encoded by a color scale
shown by the bars. The various blocks in \eqref{Hblock} are separated by grey lines.
As expected, the dominant (large negative) entries correspond to nearest-neighbour hopping
in the diagonal blocks. These are accompanied by subdominant matrix elements corresponding
to long-range hopping within each segment, reminiscent to the structure observed for
the single interval case \cite{Eisler/Peschel17}.

%%%%%%%%%%%%%%%%%%%%%%%%%%% Fig.13 %%%%%%%%%%%%%%%%%%%%%
%%%%%%%%%%%%%%%%%%%%%%%%%%%%%%%%%%%%%%%%%%%%%%%%%%%%%%%
\clearpage

The novel feature is the structure in the off-diagonal blocks,
which was visible in Ref.\,\cite{Arias_etal17_1} but not further studied there. According to the continuum results
of Sec.\,\ref{sec:cont}, the off-diagonal block should contain only single hopping terms between the
conjugate sites, which lie on a hyperbola as indicated by the dashed black lines in Fig.\,\ref{fig:EHmatrix}.
On the lattice one obtains a more complicated structure, however, one can still recognize
the location of the hyperbolae from the color-coded plots. To better visualize the structure of
the off-diagonal block, we also give 3D-plots on the right of Fig.\,\ref{fig:EHmatrix},
showing the absolute values of the matrix elements. Here the hyperbolic shape is even more
evident, one has, however, an additional structure along the antidiagonal as the distance
between the segments decreases. For large distances the hyperbola moves towards
the diagonal and the structure becomes increasingly more peaked with decreasing amplitude.

One should emphasise that the dominant amplitudes in the diagonal blocks are two orders of magnitude
larger than those in the off-diagonal ones. Furthermore, the former show an extensive scaling with $N$,
as one would expect from the local weight function \eqref{beta-loc-2int-sym}, which has a dimension of length.
In contrast, the matrix elements in the off-diagonal block do not scale with $N$, as the corresponding
bi-local weight \eqref{beta-bi-loc-2int-sym} is dimensionless. However, in order to recover the functions
$\betax$ and $\betabix$ from the lattice data, one needs some further considerations.
Indeed, it was already argued in Ref.\,\cite{Arias_etal17_1} that the longer-range hopping terms on the lattice
should be included properly to recover the local weight function. Moreover, for a single interval
the continuum limit can even be carried out analytically, recovering the CFT expression \cite{Eisler/Tonni/Peschel19}.
In the following we shall demonstrate how the continuum limit works for the double interval,
both for the local as well as bi-local weights.

\section{Continuum limit}
\label{sec_cont_2int}

The continuum limit provides a relation between the entanglement Hamiltonian on the lattice and the one obtained for the Dirac fermion theory.
It amounts to introduce a lattice spacing $s \to 0$ and consider the thermodynamic limit $N_{\sigma}\to \infty$
such that $N_{\sigma} s=\ell_{\sigma}$ is fixed, with $\sigma=1,2$.
Due to the block structure \eqref{Hblock} for the double interval, one should distinguish between the diagonal and
off-diagonal blocks of the entanglement Hamiltonian. As expected, the former one shall reproduce the local kinetic term \eqref{general},
whereas the latter one corresponds to the bi-local contribution \eqref{H_A-bi-local-def}.

We first consider the diagonal blocks and proceed similarly as in the case of a single interval \cite{Eisler/Tonni/Peschel19}.
Namely, we rewrite the entanglement Hamiltonian as an inhomogeneous long-range hopping model
\eq{
\mathcal{H}^{ \textrm{\tiny $( \sigma )$}} =
-\sum_{i \in A_\sigma} t_0(i) \,c_i^\dagger c_i 
-\sum_{i \in A_\sigma} \sum_{r\ge 1} t_r(i+r/2)
\Big( 
c_i^\dagger c_{i+r} + c_{i+r}^\dagger c_i 
\Big)\,,
\label{EHdiag}
}
where the hopping amplitude satisfies
\eq{
t_r(i+r/2)=
\Bigg\{\,
\begin{array}{ll}
-H^{ \textrm{\tiny $( \sigma )$}}_{i,i+r} \hspace{.6cm} & i, i+r \in A_\sigma  \,,
\\
0 & \mathrm{otherwise}\,.
\end{array}
\label{tri}
}
Note that the argument of the $r$-th neighbour amplitude corresponds to the midpoint of the sites
involved and we assume it to vary slowly in the variable $i$. We then introduce the continuous
coordinate $x=i \, s$, and apply the usual substitution for the fermion operators
\eq{
c_i  \;\longrightarrow\;
\sqrt{s} \,\Big(  e^{\textrm{i} q_F x}\, \psi_\textrm{\tiny R}(x) +  e^{-\textrm{i} q_F x}\,  \psi_\textrm{\tiny L}(x) \Big)\,,
\label{cpsi}}
thus rewriting them in terms of right- and left-moving fields. The phase factors are needed to
account for the quick oscillations on the lattice. In turn, the substitution \eqref{cpsi} amounts
to linearizing the dispersion and shifting the Fermi points to $q=0$, thus reproducing the
relativistic dispersion.

Carrying out the continuum limit is then rather straightforward. Substituting \eqref{cpsi} into
\eqref{EHdiag}, Taylor expanding the fields as well as the hopping amplitude
$t_r(i+r/2) \to t_r(x)+r \, s \, t_r'(x)/2$ to first order in $s$, and dropping oscillatory terms,
one arrives at the following expression \cite{Eisler/Tonni/Peschel19}
\eq{
\mathcal{H}^{ \textrm{\tiny $( \sigma )$}}
 \;\longrightarrow\;
%\int_{a_\sigma}^{b_\sigma} dx \,  F_1(x)\,
%T_{00}(x)
%-\int_{a_\sigma}^{b_\sigma} dx \, F_0(x)\, 
%N(x)
\int_{A_\sigma} \textrm{d} x \, \Big[ v(x) \, T_{00}(x) - \mu(x) \, N(x) \Big]\,,
\label{EHdcont}}
where $T_{00}(x)$ is given by \eqref{T00-def} and $N(x)$ is the number operator
\eq{
N(x)=
\psi^\dagger_\textrm{\tiny R}(x) \, \psi_\textrm{\tiny R}(x) + \psi^\dagger_\textrm{\tiny L}(x) \, \psi_\textrm{\tiny L}(x)\,,
}
while the functions $\mu(x) $ and $v(x)$ are defined as 
\eq{
v(x) \equiv\, 2s  \sum_{r=1}^{\infty}  r \,\sin(r q_F s) \,t_r(x) \,,
\; \qquad \;
\mu(x) \equiv t_0(x) + 2 \sum_{r=1}^{\infty}  \cos(r q_F s) \,t_r(x) \,.
\label{vmu}}
The result \eqref{EHdcont} is an inhomogeneous Dirac theory with velocity parameter $v(x)$
and a chemical potential $\mu(x)$. Hence, contrary to naive expectations, the Fermi velocity
in the continuum does not simply correspond to the nearest-neighbour hopping on the lattice,
but is rather modified by the presence of long-range hopping. Furthermore, in order to get
a finite result for the velocity $v(x)$, the hopping terms $t_r(x)$ should scale with the size
$N_\sigma$ of the segment, which is indeed the case as pointed out in sec. \ref{sec:lattice}.
In contrast to the single interval case \cite{Eisler/Peschel17}, however, we have no analytical results on $t_r(x)$
and the sums in \eqref{vmu} must be evaluated numerically. In particular, to recover the local piece
$\mathcal{H}_{\textrm{\tiny loc}}=\mathcal{H}^{\textrm{\tiny (1)}}+\mathcal{H}^{\textrm{\tiny (2)}}$
one needs to find $v(x)=2\pi \beta(x)$ and $\mu(x)=0$.

Let us now consider the off-diagonal blocks in \eqref{Hblock}. Using the symmetry of the matrix,
one could define the corresponding piece of the entanglement Hamiltonian as
\eq{
\mathcal{H}^{\textrm{\tiny (1,2)}} \,=\, 
\frac{1}{2}  
\sum_{\substack{i \in A_1 \\  j \in A_2}}  
%\sum_{ i \in A_1 , j\in A_2} 
\!H^{\textrm{\tiny (1,2)}}_{i,j} \Big(c^{\dag}_i c_j + c^{\dag}_j c_i\Big) \, .
\label{EH12}}
We would like to reproduce the bi-local term \eqref{H_A-bi-local-def} by an appropriate continuum limit,
we thus fix $i s=x$ and look for the contributions around the conjugate site $js \approx x_{\textrm{\tiny c}}$.
Since the matrix elements in the off-diagonal block do not scale with the segment size,
it is enough to keep the zeroth order term in the expansion of the fields.
This is given by
\bea
c_i^\dagger c_{j} + c^{\dag}_j c_i
%\textrm{h.c.}
&\longrightarrow&
s \; \Big[\,
\ex^{\textrm{i} q_F s(j-i)}  \psi^\dagger_\textrm{\tiny R}(x) \, \psi_\textrm{\tiny R}(x_{\textrm{\tiny c}}) +
\ex^{-\textrm{i} q_F s(j-i)}  \psi^\dagger_\textrm{\tiny L}(x) \, \psi_\textrm{\tiny L}(x_{\textrm{\tiny c}})  
%+ \textrm{h.c.} 
\nonumber
\\
& & \hspace{.2cm}
+\,
\ex^{\textrm{i} q_F s(j+i)}  \psi^\dagger_\textrm{\tiny L}(x) \, \psi_\textrm{\tiny R}(x_{\textrm{\tiny c}}) +
\ex^{-\textrm{i} q_F s(j+i)}  \psi^\dagger_\textrm{\tiny R}(x) \, \psi_\textrm{\tiny L}(x_{\textrm{\tiny c}})  
+ \textrm{h.c.} \,\Big] \,.
\label{bilochop}
\eea
Note that in the above expression each term has an oscillatory factor with a large argument.
In fact, it is not clear a priori, which one of these term provides a proper continuum limit,
i.e. a smooth function of $i$. Since for the infinite chain the left- and
right-moving fermions are not supposed to mix, we will consider the exponentials with the
$j-i$ factors and will provide further justification later. The bi-local piece of the entanglement Hamiltonian then reads
\eq{
\mathcal{H}^{\textrm{\tiny (1,2)}} 
\;\longrightarrow\;
\int_{A_1} \dd x \,
\Big[ \, \mathcal{S}(x) \,  T_{\textrm{\tiny bi-loc}}(x,x_{\textrm{\tiny c}})+ \mathcal{C}(x) \, \tilde T_{\textrm{\tiny bi-loc}}(x,x_{\textrm{\tiny c}}) \, \Big] \,,
\label{EH12cont}}
where $T_{\textrm{\tiny bi-loc}}(x,x_{\textrm{\tiny c}})$ is the operator defined in \eqref{T-bilocal-2int} and
\eq{
\tilde T_{\textrm{\tiny bi-loc}}(x,x_{\textrm{\tiny c}} ) 
= 
\frac{1}{2} \left[\,
\psi^\dagger_\textrm{\tiny R}(x) \, \psi_\textrm{\tiny R}(x_{\textrm{\tiny c}}) +
\psi^\dagger_\textrm{\tiny L}(x) \, \psi_\textrm{\tiny L}(x_{\textrm{\tiny c}})  +\textrm{h.c.} 
\,\right] .
}
The corresponding weight functions with $x_i=i \, s$ are given by
\eq{
\mathcal{S}(x_i) = \sum_{j \in A_2} \sin[q_Fs(j-i)] \, H^{\textrm{\tiny (1,2)}}_{i,j} \, , 
\;\; \qquad \;\;
\mathcal{C}(x_i) = \sum_{j \in A_2} \cos[q_Fs(j-i)] \, H^{\textrm{\tiny (1,2)}}_{i,j} \, .
\label{scx}}
The same procedure can be carried out for the piece $\mathcal{H}^{\textrm{\tiny (2,1)}}$
with indices $i$ and $j$ in \eqref{EH12} interchanged. The continuum limit is the same
as \eqref{EH12cont} with the integral running over $A_2$, and the sums in \eqref{scx}
running over $A_1$.

\subsection{Equal intervals at half filling\label{sec:equalint}}

We start with the simplest case of equal intervals $N_1=N_2=N$ at half filling $q_F \, s=\pi/2$,
setting $A_2=\left[d+1,d+N\right]$ and $A_1=\left[-(d+N),-(d+1)\right]$ such that the distance
between the intervals is $2d+1$. 
The coordinates of the endpoints in the continuum description must be chosen 
as $a=ds$ and $b=(d+N)s$, such that $b-a=\ell$.
Due to particle-hole symmetry, $H$ has a checkerboard structure with $H_{i,i+2p}= 0$,
which immediately yields $\mu(x) = 0$, whereas $v(x)$ has to be determined numerically.
Due to the extensive scaling of the
diagonal blocks, it is useful to introduce the rescaled hopping
\eq{
h_{i-p,i+p+1} =  -H_{i-p,i+p+1}/N = \, t_{2p+1}(i)/N  \,.
}
Using this in \eqref{vmu} with $x_i=i \, s$ and fixing the scale as $Ns =\ell=1$, the velocity reads
\eq{
%\frac{F(x)}{2\pi\ell} 
v(x_i)= %\frac{1}{\pi}\,
2\sum_{p=0}^P (-1)^p \,(2p+1) \,h_{i-p,i+p+1} \,,
%^{\mathrm{(\sigma})}
\label{vxi}  
}
where we introduced a cutoff $P<N/2$. Note that for a fixed $i \in \left[d+1,d+N-1\right]$
the sums are carried out perpendicular to the main diagonal of the matrix.

\begin{figure}[t!]
\center
\includegraphics[width=0.49\textwidth]{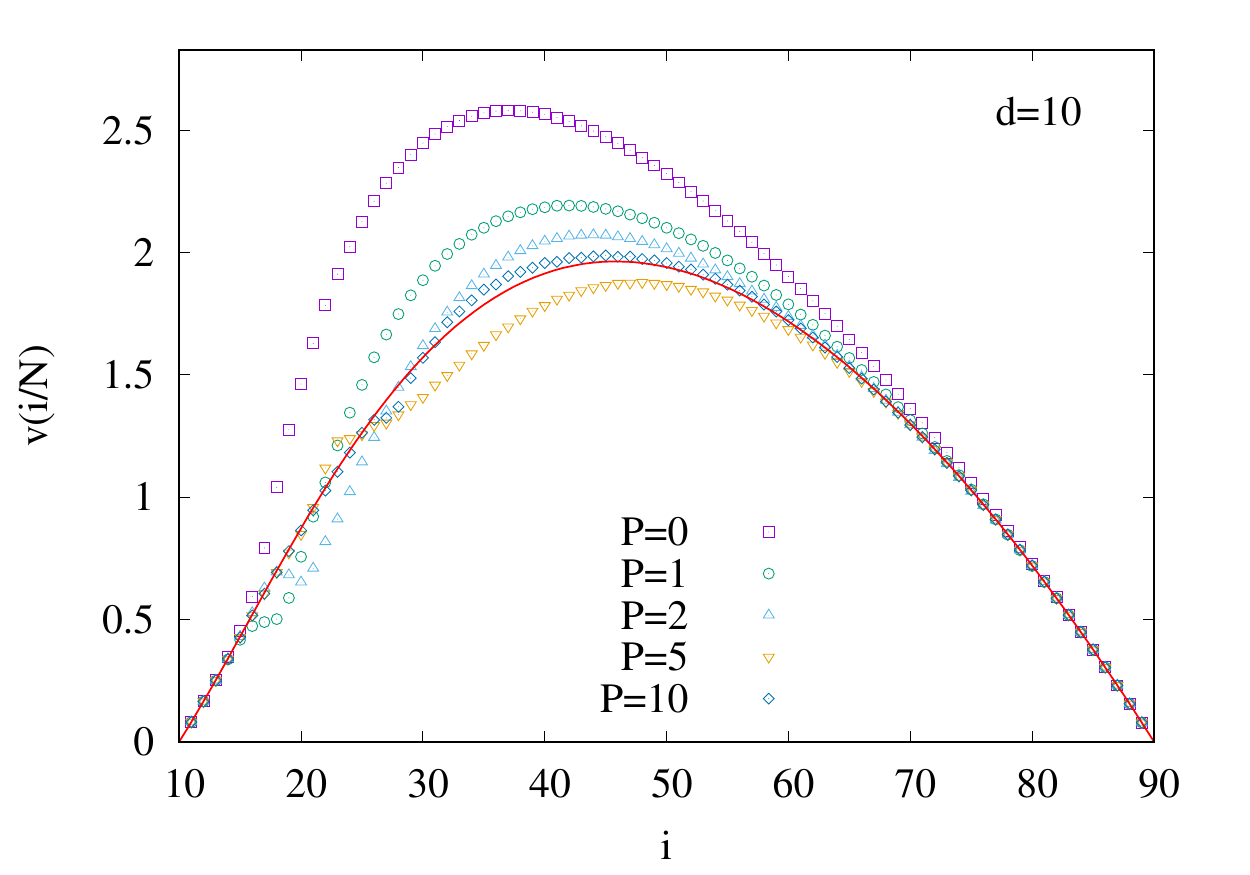}
\includegraphics[width=0.49\textwidth]{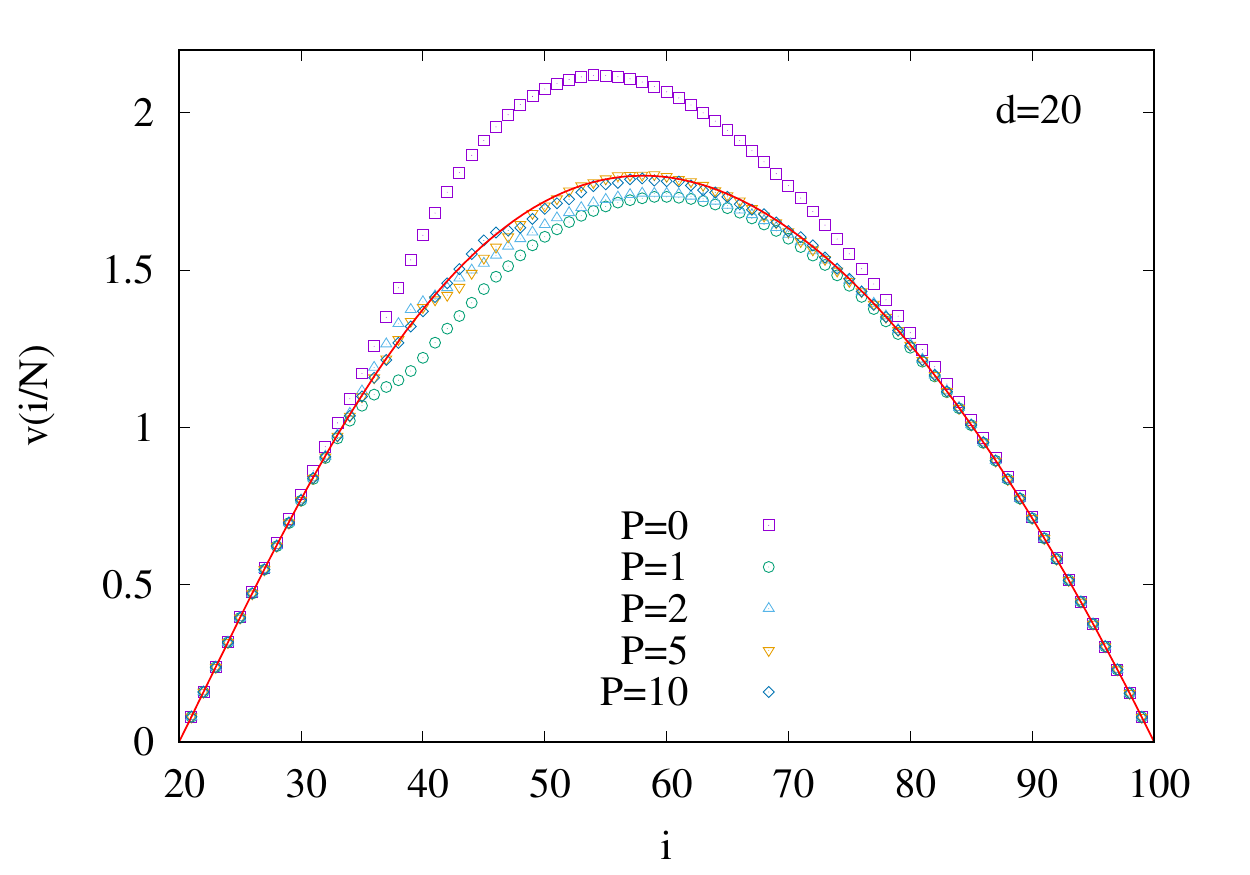}
\caption{Convergence of the velocity parameter \eqref{vxi} in the continuum
limit of the entanglement Hamiltonian for a double interval with $N=80$ and $d=10,20$.
The red solid line shows the function $2\pi\beta(i/N)$ from \eqref{beta-loc-2int-sym} setting $a=d/N$ and $b=1+d/N$.
Note that the figures always show the right segment.}
\label{fig:loc_hf}
\end{figure}

The convergence of $v(x_i)$ as a function of the cutoff is shown in Fig.\,\ref{fig:loc_hf}
for $N=80$ and $d=10,20$. The case $P=0$ is simply the scaled nearest-neighbour
hopping and one observes that as the distance between the segments becomes smaller, it deviates
more and more from the continuum limit prediction $2\pi \beta(i/N)$, shown by the red lines.
Indeed, the data for $P=0$ lies well above the red line with its maximum shifted towards the center
of the chain, and a good overlap is found only close to the boundaries. Increasing the
value of $P$ one finds a slow convergence towards the continuum limit and already for $P=10$
one has an almost perfect overlap.

%%%%%%%%%%%%%%%%%%%%%%%%%%%%%%%%%%%%%%%%%%%%%%%%%%%%%%%%%%%
%
%
%%%%%%%%%%%%%%%%%%%%%%%%%%%%%%%%%%%%%%%%%%%%%%%%%%%%%%%%%%%

%%%%%%%%%%%%%%%%%%%%%%%%%%%%%%%%%%%%%%%%%%%%%%%%%%%%%%%%%%%
%
\begin{figure}[t!]
\center
\includegraphics[width=0.49\textwidth]{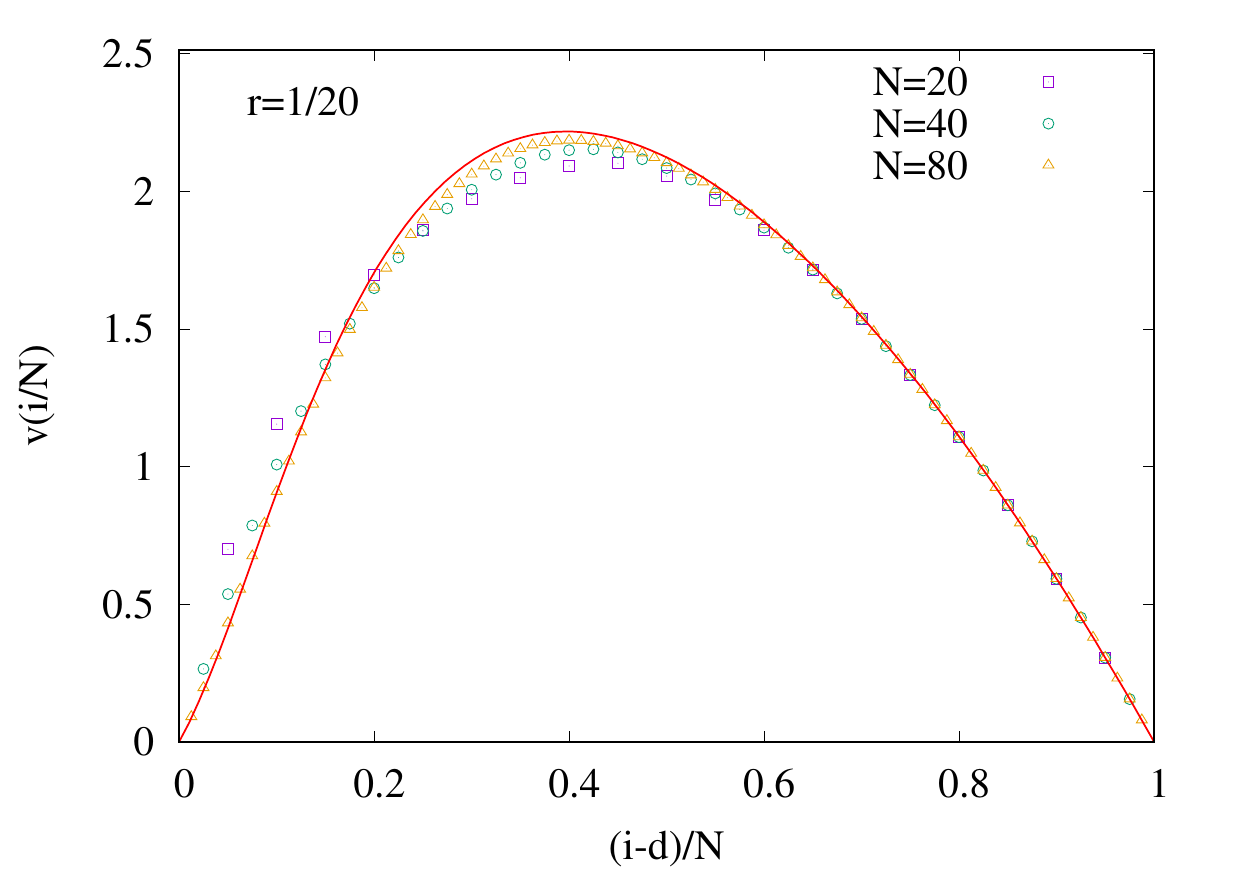}
\includegraphics[width=0.49\textwidth]{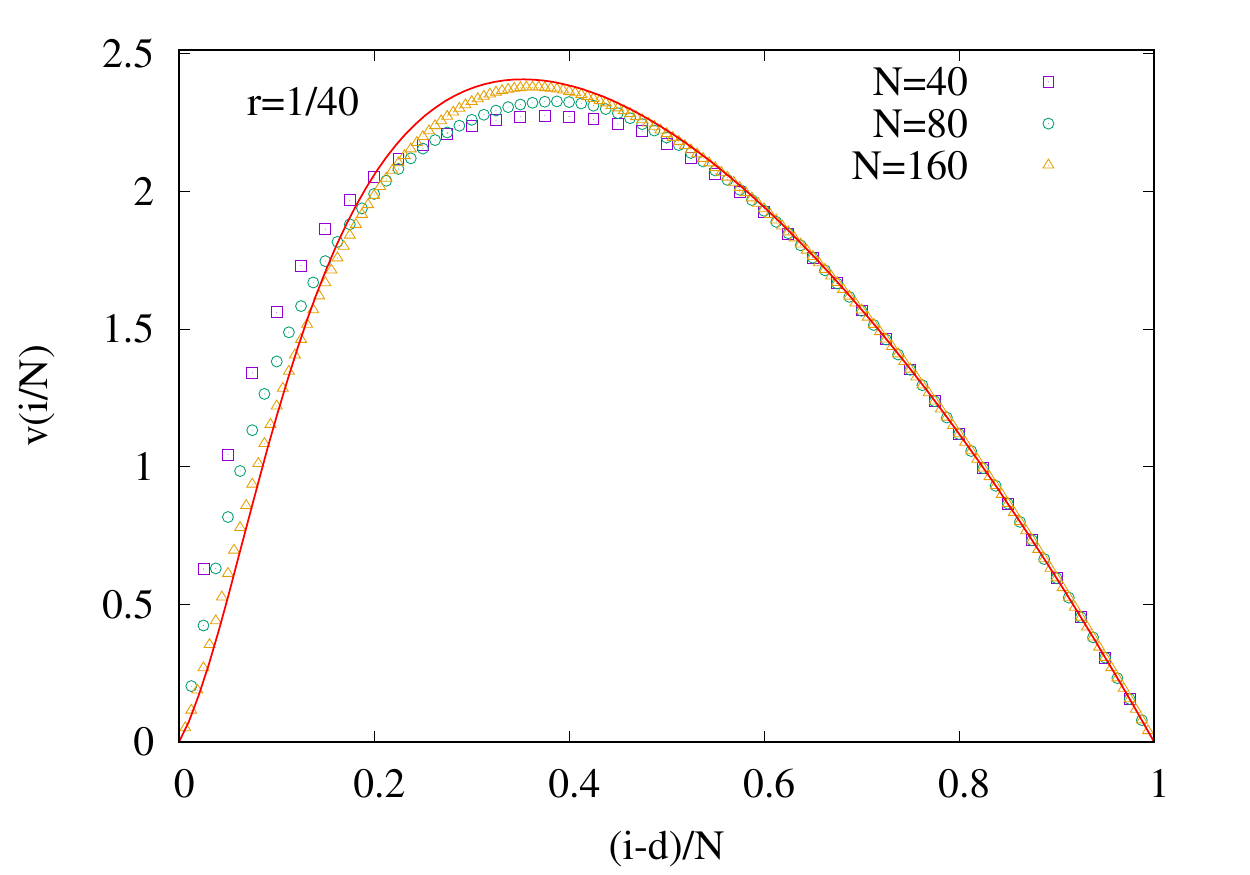}
\caption{Convergence of $v(x_i)$ %from \eqref{vxi} with $P=N/2-1$,
for a fixed ratio $r=d/N$ and increasing values of $N$, plotted against $(i-d)/N$.
The red solid lines show the weight function $2\pi\beta(r+y)$ in \eqref{beta-loc-2int-sym} as a function of $y$,
setting $a=r$ and $b=1+r$.}
\label{fig:loc_hf2}
\end{figure}
%
%%%%%%%%%%%%%%%%%%%%%%%%%%%%%%%%%%%%%%%%%%%%%%%%%%%%%%%%%%%

In the above examples the ratio $r=d/N$ between the distance and segment size is large
enough to ensure a very good convergence already for moderate $N$. The situation becomes more
complicated for small ratios $r$, as shown in Fig.\,\ref{fig:loc_hf2}. Here the velocity $v(x_i)$
is calculated using the maximal cutoff $P=N/2-1$ for various $d$ and $N$, keeping their ratio fixed.
One finds that for $r=1/20$ already $N=80$ is sufficient to converge the data, whereas for
$r=1/40$ the data set $N=160$ still shows some visible deviations. Obviously, this is because
the number of sites $2d+1=9$ between the segments is still too small to ensure a reasonable
continuum limit. Nevertheless, Fig.\,\ref{fig:loc_hf2} convincingly demonstrates that the limit
is approached smoothly for $N\to \infty$.

\begin{figure}[t!]
\center
\includegraphics[width=0.49\textwidth]{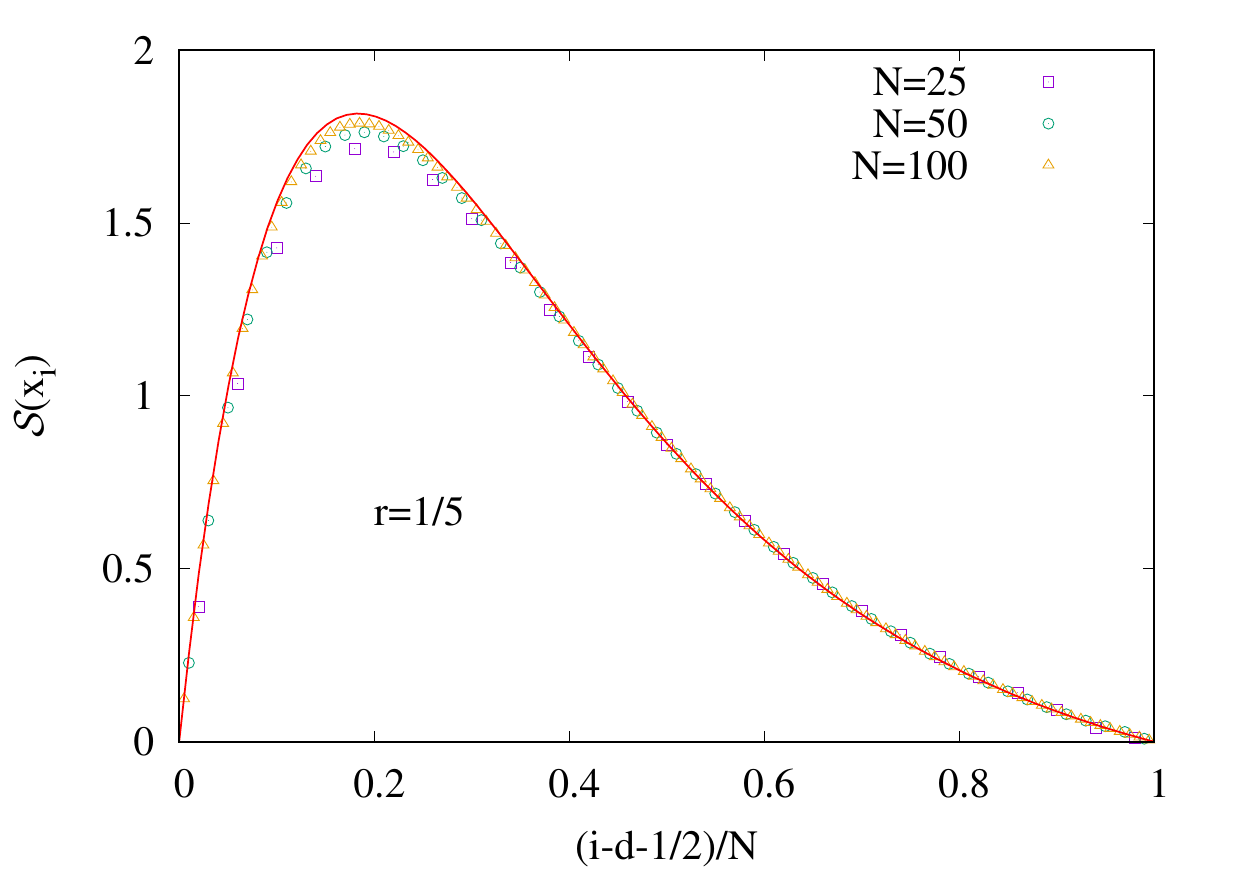}
\includegraphics[width=0.49\textwidth]{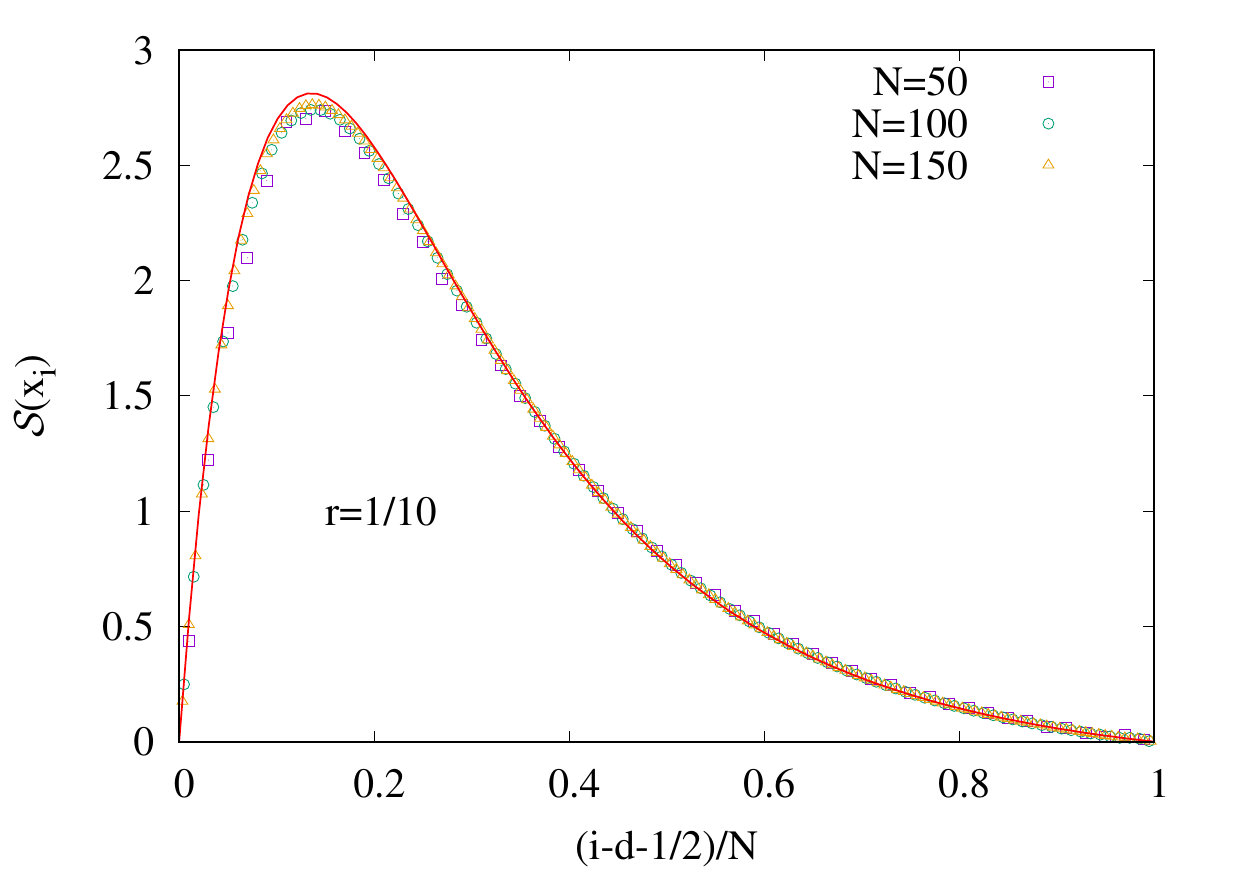}
\caption{Bi-local weight function $\mathcal{S}(x_i)$ at half filling for increasing segment sizes $N$ and two different ratios $r=d/N$,
shown against the variable $x_i-r$. The red lines show the continuum result $2\pi\betabix$ for comparison.}
\label{fig:biloc_hf}
\end{figure}

Finally, we consider the off-diagonal scaling functions in \eqref{scx}.
Note that here the range of the lattice variable is $i\in\left[d+1,d+N\right]$ and we chose
$a=ds$ and $b=(d+N)s$ for the boundaries in the continuum case. Since the weight function must vanish
exactly at $x=b$, it is useful to slightly shift the discretized coordinates $x_i=(i-1/2)s$ to get a better
overlap with the continuum results. 
At half filling the function $\mathcal{C}(x_i)=0$ vanishes identically, while $\mathcal{S}(x_i)$ simplifies to
\eq{
\mathcal{S}(x_i) = \sum_{j \in A_1} (-1)^{(j-i-1)/2} H_{i,j} \, , 
\;\;\;\; \qquad \;\;\;\;
x_i \in (a,b) \, .
\label{sxihf}}
In contrast to the local term, this definition involves an alternating row-wise
sum of the matrix elements. The result is shown in Fig.\,\ref{fig:biloc_hf} for increasing segment sizes $N$ and
fixed ratios $r=1/5$ and $r=1/10$. In both cases one obtains a very good convergence towards the
bi-local weight $2\pi \betabix$, shown by the red lines. This function is more concentrated near the left end of the
segment than $2 \pi \beta(x)$ due to the factor $1/x$ in \eqref{beta-bi-loc-2int-sym}, and
the effect increases for smaller distances.
Note that the relation $\mathcal{S}(-x_i)=-\,\mathcal{S}(x_i)$ follows
by symmetry arguments. One should also remark that keeping the terms in \eqref{bilochop} with $i+j$
would lead to another sum that is simply related to \eqref{sxihf} by $(-1)^i \mathcal{S}(x_i)$. Obviously,
this is an alternating expression which vanishes in the continuum limit, thus justifying our choice
of dropping these terms.

%%%%%%%%%%%%%%%%%%%%%%%%%%%%%%%%%%%%%%%%%%%%%%%%%%%%%%%%%%%
%
%
%%%%%%%%%%%%%%%%%%%%%%%%%%%%%%%%%%%%%%%%%%%%%%%%%%%%%%%%%%%

\subsection{Arbitrary filling}

We now move to the case of arbitrary filling, choosing equal intervals for simplicity.
For a single interval one can actually show analytically, that the continuum limit leads to the exact same
expressions as in the half-filled case \cite{Eisler/Tonni/Peschel19}. This proof uses the analytical expressions for $t_r(x)$,
which are nonzero for both even and odd $r$ \cite{Eisler/Peschel17}, to evaluate the sums in \eqref{vmu}. In the
present case, where the sums are carried out numerically, this leads to some complications
as the hopping profiles \eqref{tri} on the lattice are associated to integer and half-integer sites for
even and odd $r$, respectively. Thus, analogously to the odd sums as in \eqref{vxi} for the half-filled
case, one could define an even sum, albeit with coordinates assigned as $x_i=(i-1/2)s$. The even
and odd sums can then be added by summing the $i$-th term of the odd sequence with the average
of the $i$-th and $(i+1)$-th terms of the even sequence. Alternatively, one could also define the row-wise
sums with $x_i=(i-1/2)s$
\eq{
%\frac{F(x)}{2\pi\ell} 
v(x_i)\,= %\frac{1}{\pi}\,
\! \sum_{j \in A_2} \sin[q_Fs(j-i)]\, (j-i) h_{i,j} \,, 
\;\;\; \qquad \;\;\;
\mu(x_i)\,=
\, -\! \sum_{j \in A_2} \cos[q_Fs(j-i)] \,H_{i,j}  \,,
%^{\mathrm{(\sigma})}
\label{vmurow}  
}
which also give only corrections of order $s^2$ that vanish in the continuum limit. Note that the factor
of two as compared to \eqref{vmu} is now missing, since the sums run over an entire row of the
matrix and not only in the upper diagonal part.

%%%%%%%%%%%%%%%%%%%%%%%%%%%%%%%%%%%%%%%%%%%%%%%%%%%%%%%%%%%
%
\begin{figure}[t!]
\center
\includegraphics[width=0.49\textwidth]{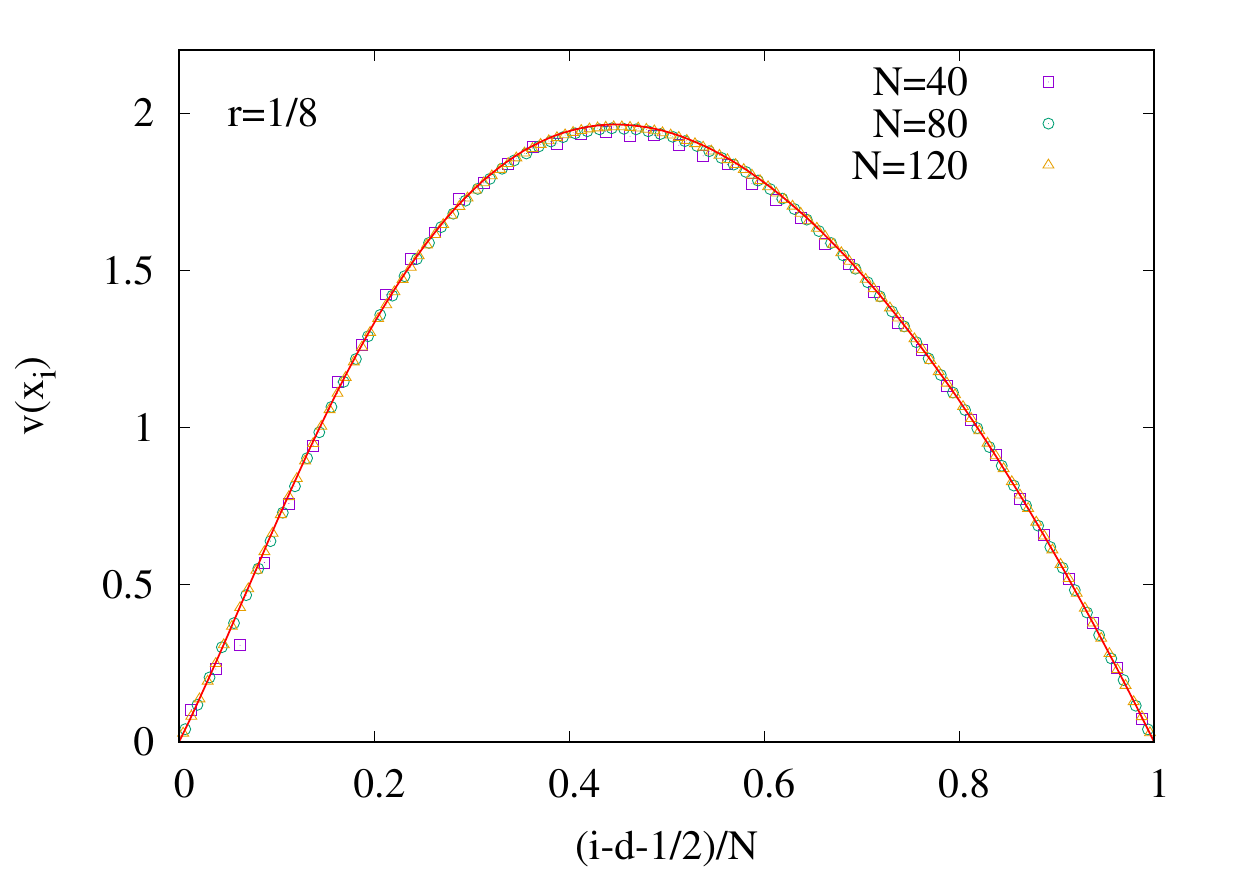}
\includegraphics[width=0.49\textwidth]{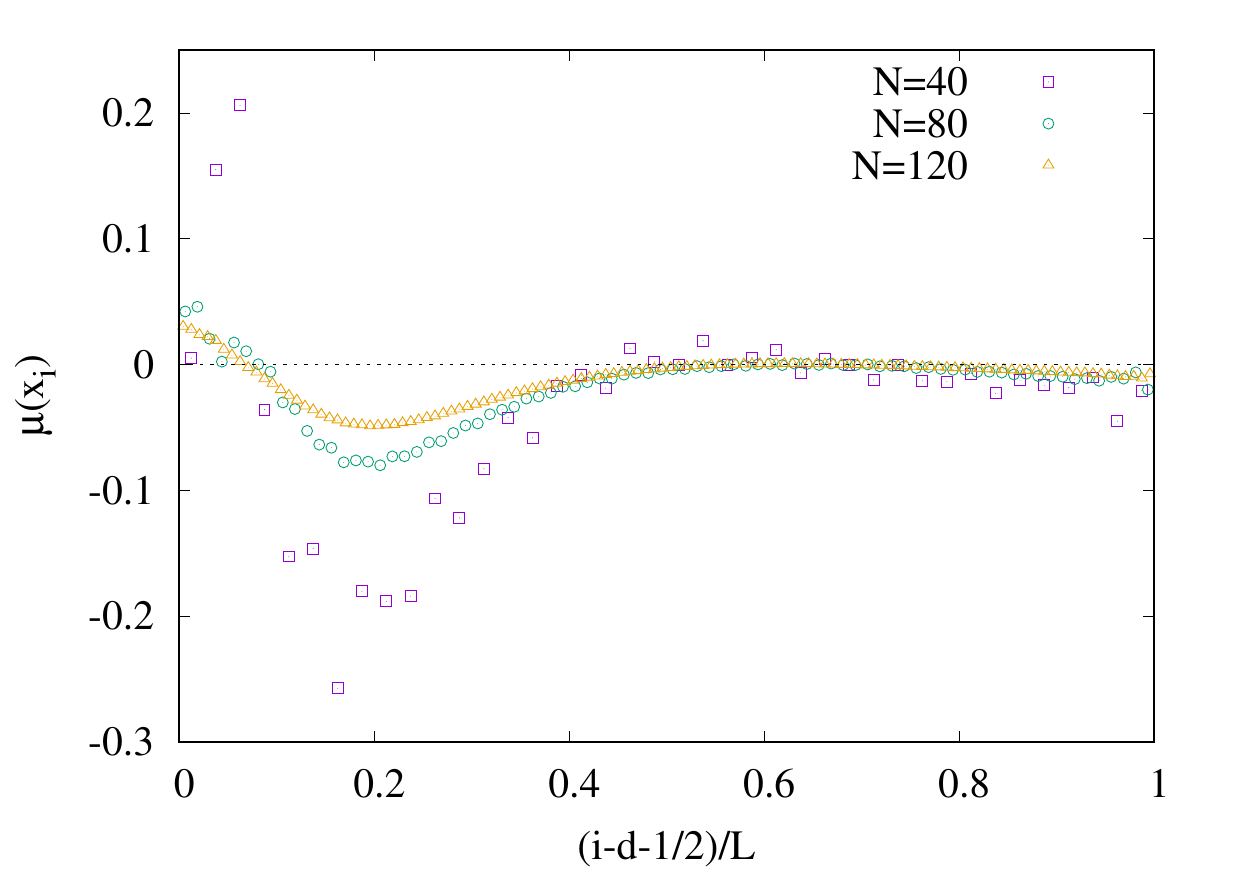}
\includegraphics[width=0.49\textwidth]{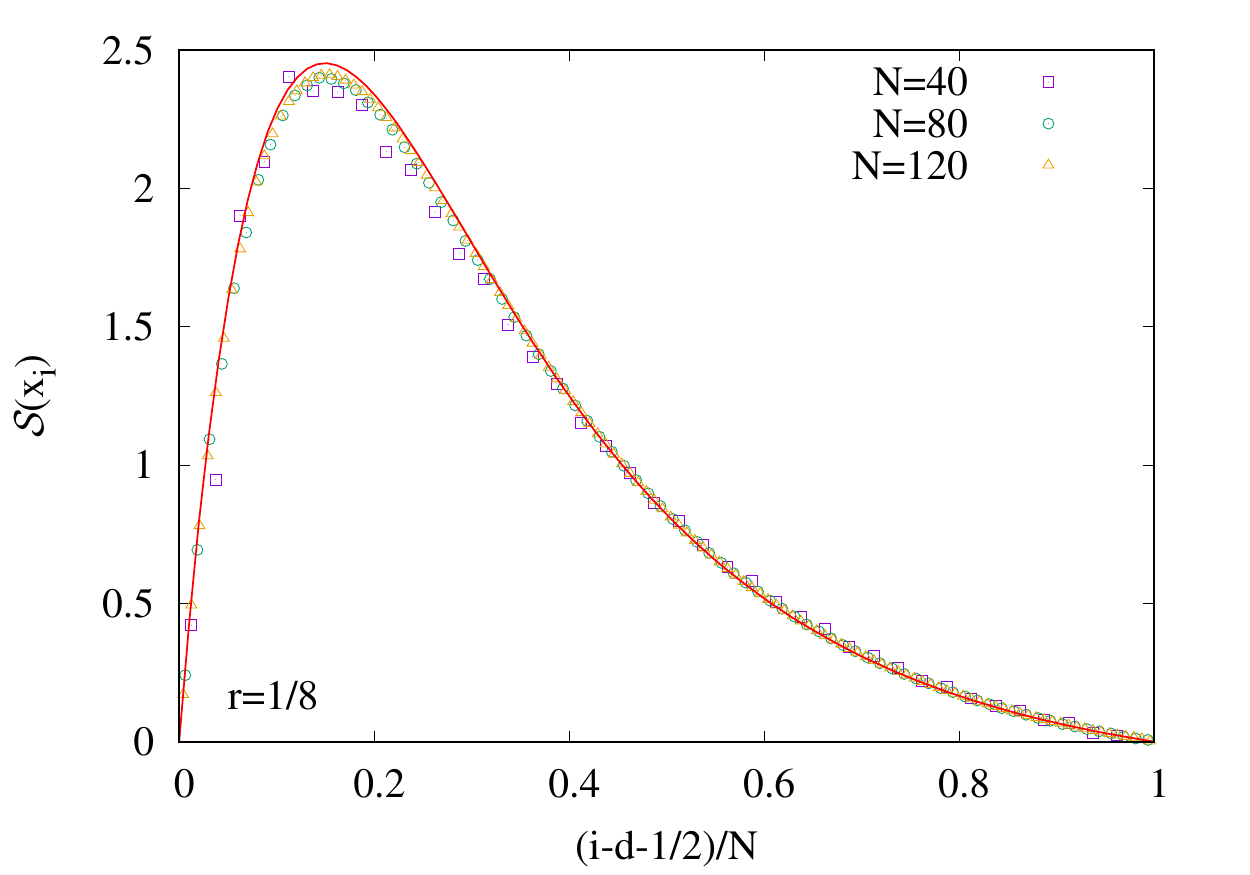}
\includegraphics[width=0.49\textwidth]{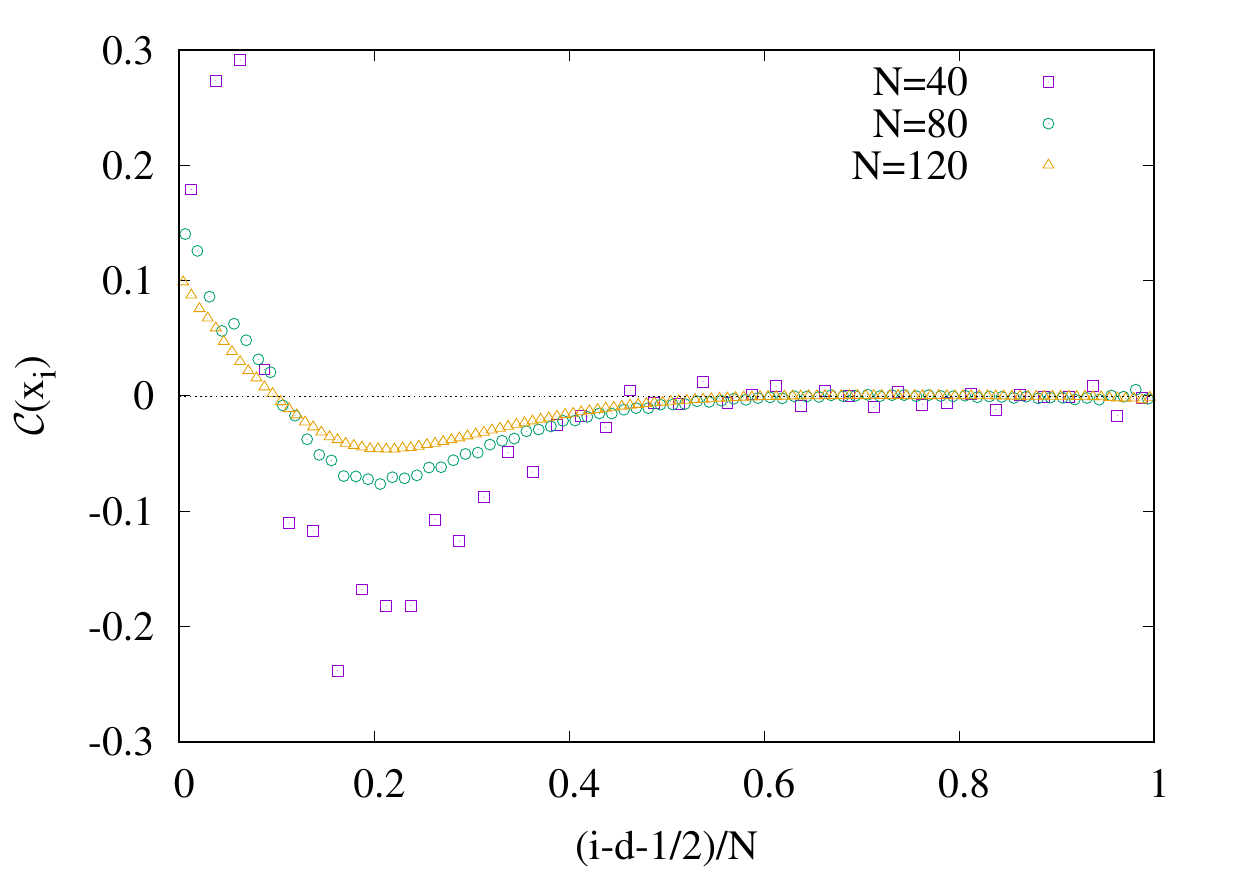}
\caption{Top panel: velocity $v(x_i)$ (left) and chemical potential $\mu(x_i)$ (right) from \eqref{vmurow} at $1/3$ filling
with a ratio $r=1/8$. The red solid line shows the result $2\pi \beta(x)$.
Bottom panel: bi-local weight functions $\mathcal{S}(x_i)$ (left) and $\mathcal{C}(x_i)$ (right) from \eqref{scx}.
The red solid line shows the result $2\pi \betabix$.}
\label{fig:contlim_fill}
\end{figure}
%
%%%%%%%%%%%%%%%%%%%%%%%%%%%%%%%%%%%%%%%%%%%%%%%%%%%%%%%%%%%

We observed that the row-wise sums yield a smoother convergence towards the expected weight functions.
This is essentially due to the large oscillations as a function of $i$ in the even/odd diagonal sums,
which do not cancel perfectly when following the averaging process described above.
The row-wise sums \eqref{vmurow} are shown in the top panel of Fig.\,\ref{fig:contlim_fill} for $1/3$ filling (corresponding to $q_Fs=\pi/3$)
for a ratio $r=1/8$ and increasing $N$. For $v(x_i)$ on the left hand side, the convergence towards $2\pi\beta(x)$
is excellent already for the moderate segment sizes used. The chemical potential on the right
shows a much slower convergence towards zero. However, this is a result of
rather nontrivial cancellations in the sums \eqref{vmurow}, since the definition of $\mu(x_i)$ involves
the unscaled $H_{i,j}$. We also show the results for the bi-local weight functions \eqref{scx},
given by row-wise sums for generic fillings, in the bottom panel of Fig.\,\ref{fig:contlim_fill}.
The comparison of $\mathcal{S}(x_i)$ against $2\pi \betabix$ shows a very good agreement,
with only tiny deviations visible around the maximum. Similarly to $\mu(x_i)$, the
bi-local weight $\mathcal{C}(x_i)$ defined with the cosine function goes towards zero as it should.

\subsection{Unequal intervals}

\begin{figure}[t!]
\center
\includegraphics[width=0.49\textwidth]{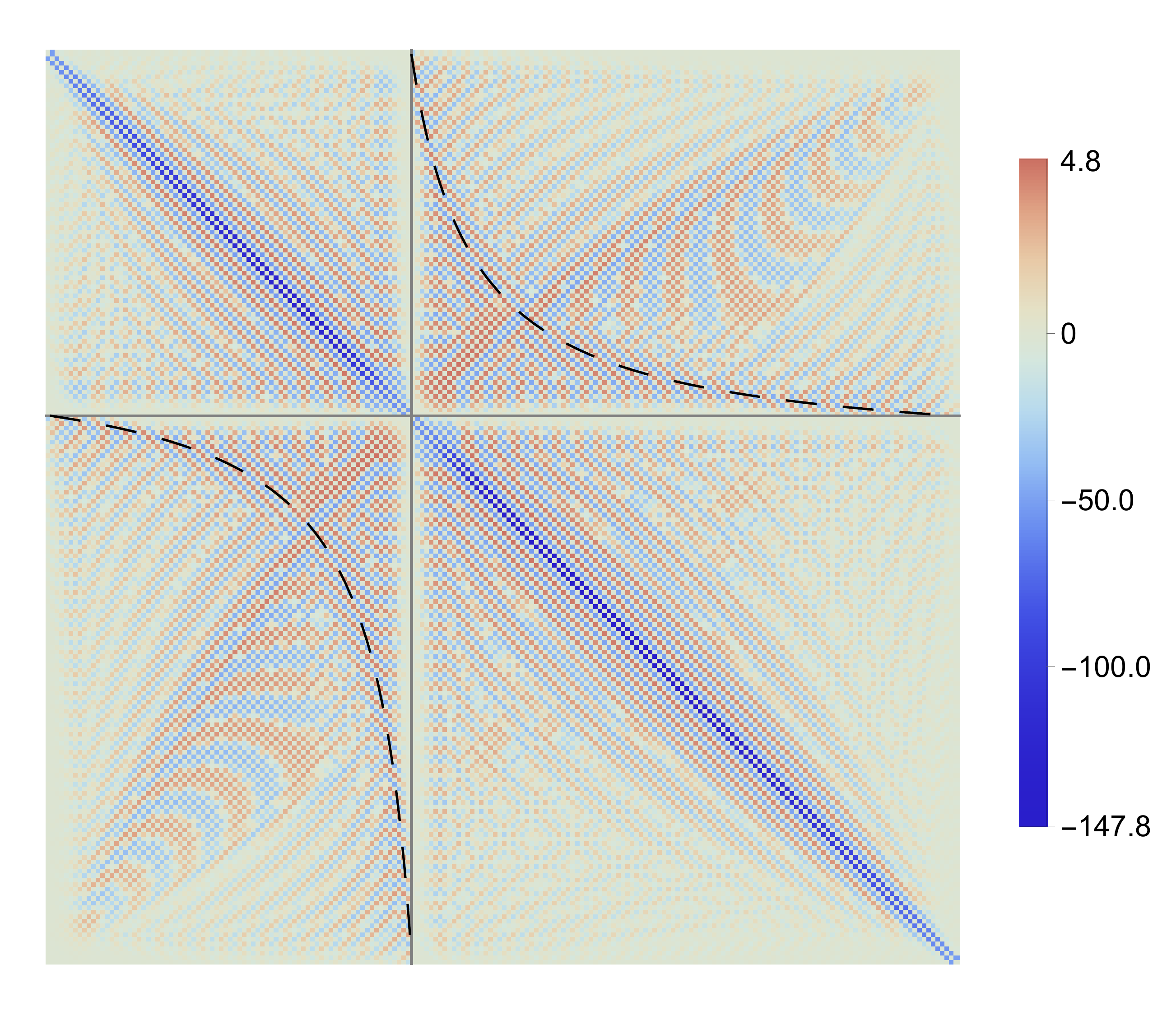}
\includegraphics[width=0.49\textwidth]{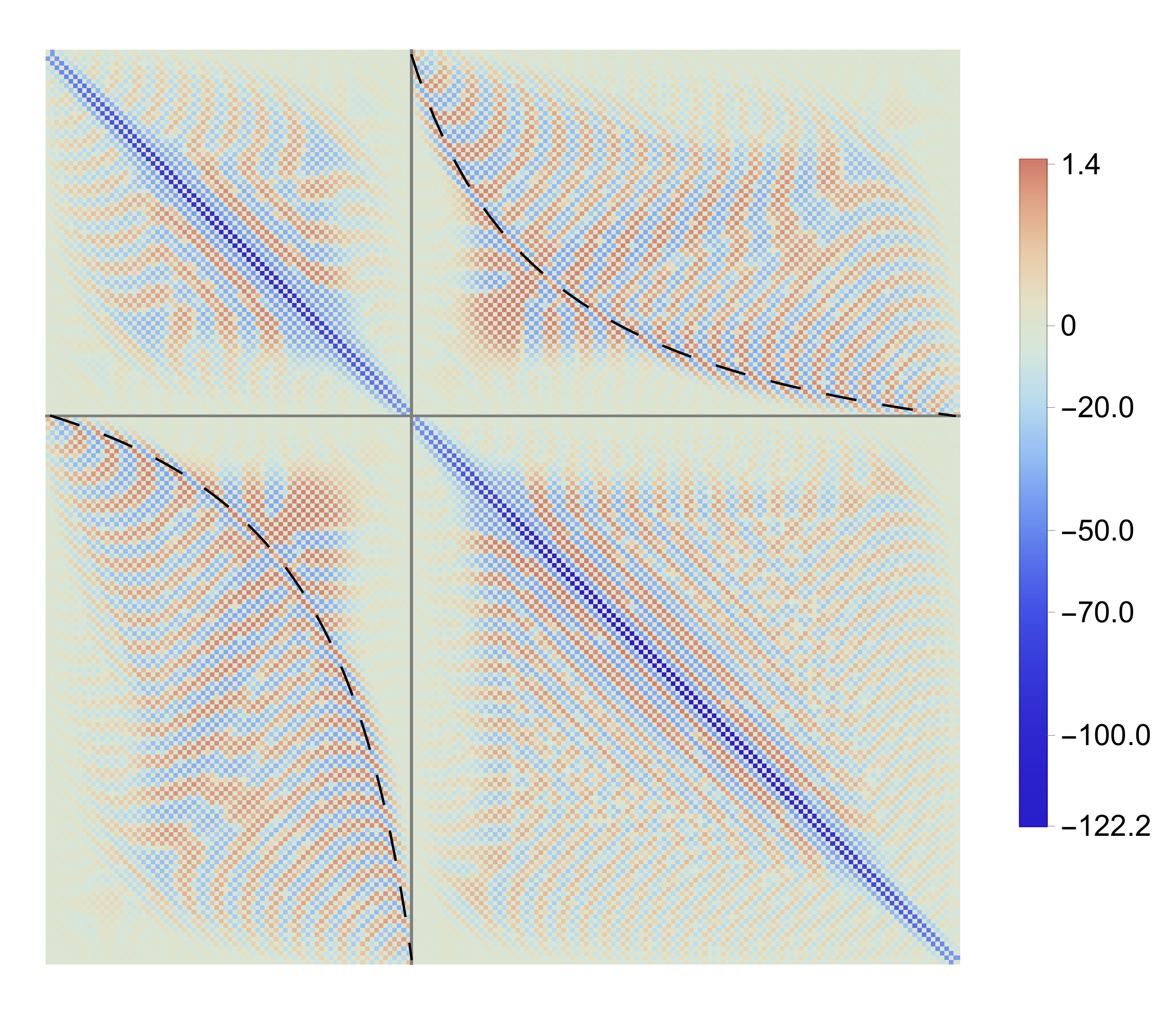}
\caption{Matrix elements in $H$ for the case of unequal segments
with $N_1=80$, $N_2=120$ and $D=21$ (left) and $51$ (right).}
\label{fig:matrix_unequal}
\end{figure}

As a final example, we discuss the case of unequal intervals in a half-filled chain.
Due to translational invariance, one can make the choice $A_1=\left[1,N_1\right]$
and $A_2=\left[N_1+D+1,N_1+D+N_2\right]$ for the intervals of size $N_1$ and $N_2$,
separated by $D$ sites. The matrix plot in Fig.\,\ref{fig:matrix_unequal} shows that
the main features are similar to the symmetric case, and the hyperbola $x_{\textrm{\tiny c}}(x)$
can again be recognized. However, the fine structure in the off-diagonal block differs
from the one in Fig.\,\ref{fig:EHmatrix}, with a curved structure appearing also along the
antidiagonal.

%%%%%%%%%%%%%%%%%%%%%%%%%%%%%%%%%%%%%%%%%%%%%%%%%%%%%%%%%%%
%
%
%%%%%%%%%%%%%%%%%%%%%%%%%%%%%%%%%%%%%%%%%%%%%%%%%%%%%%%%%%%

%%%%%%%%%%%%%%%%%%%%%%%%%%%%%%%%%%%%%%%%%%%%%%%%%%%%%%%%%%%
%
\begin{figure}[t!]
\center
\includegraphics[width=0.49\textwidth]{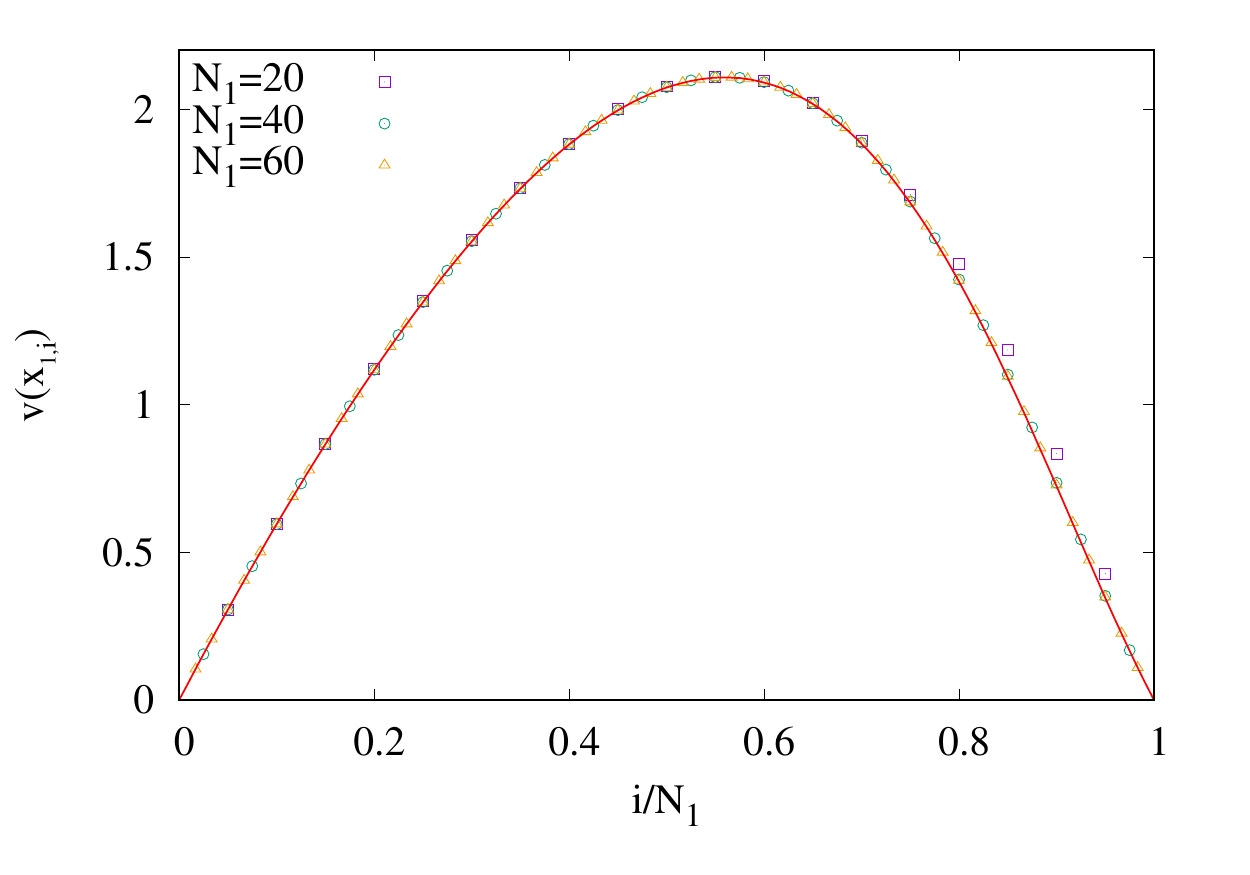}
\includegraphics[width=0.49\textwidth]{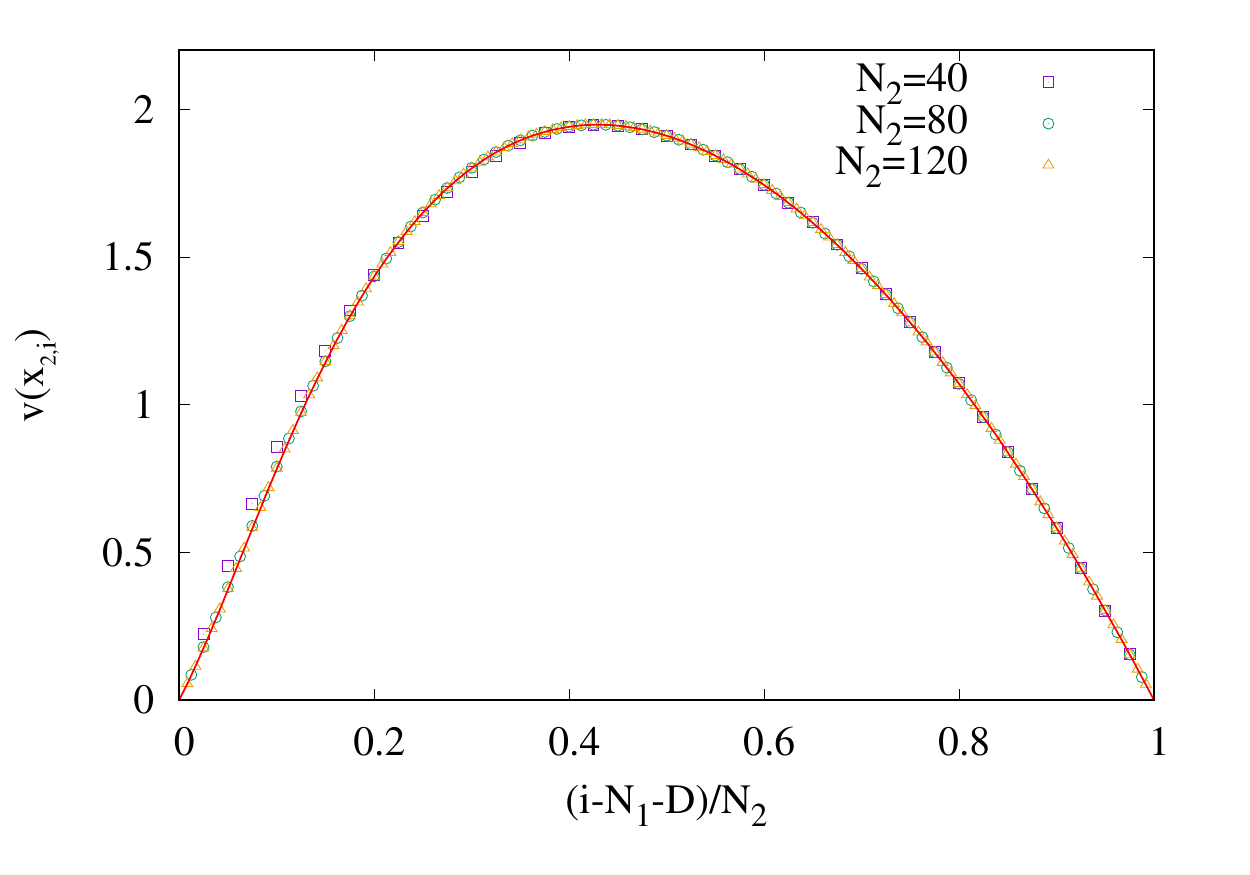}
\includegraphics[width=0.49\textwidth]{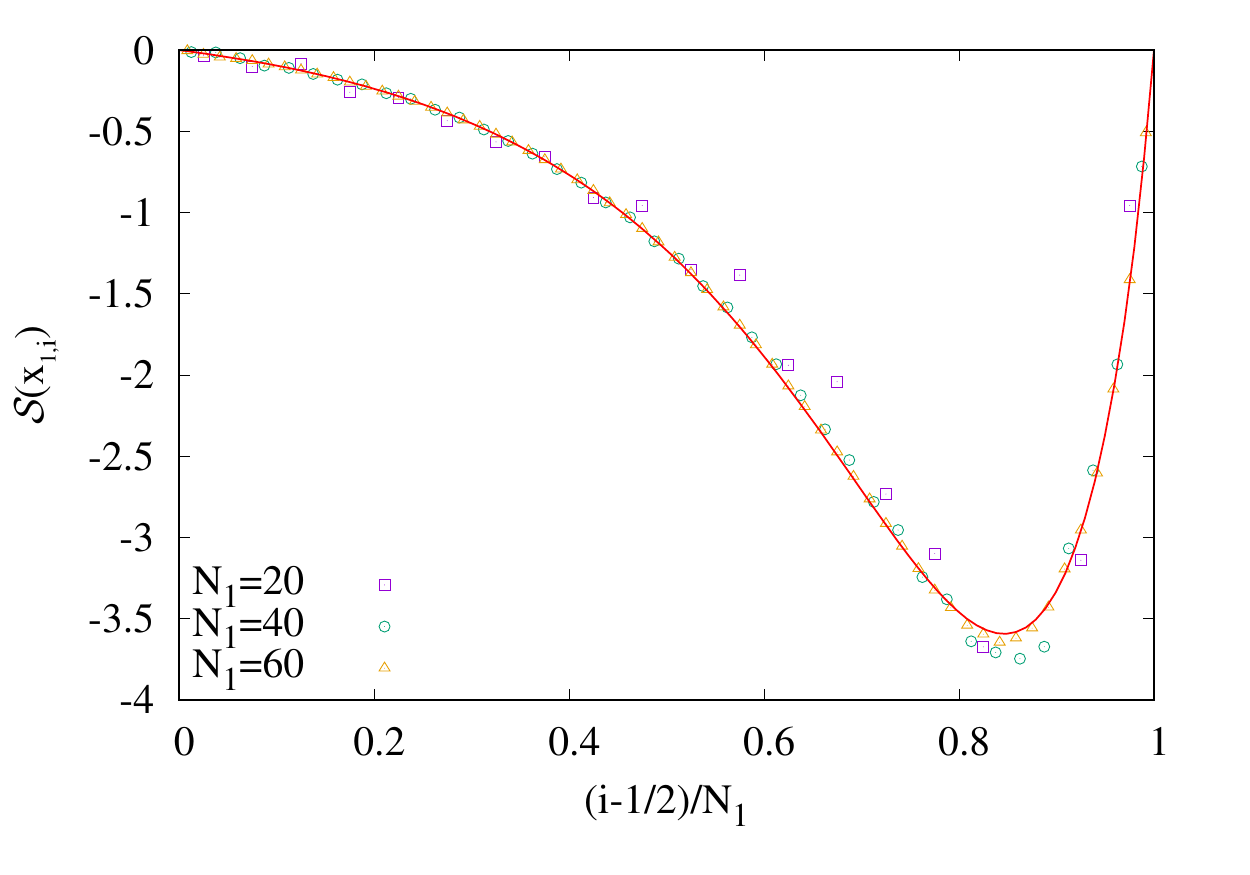}
\includegraphics[width=0.49\textwidth]{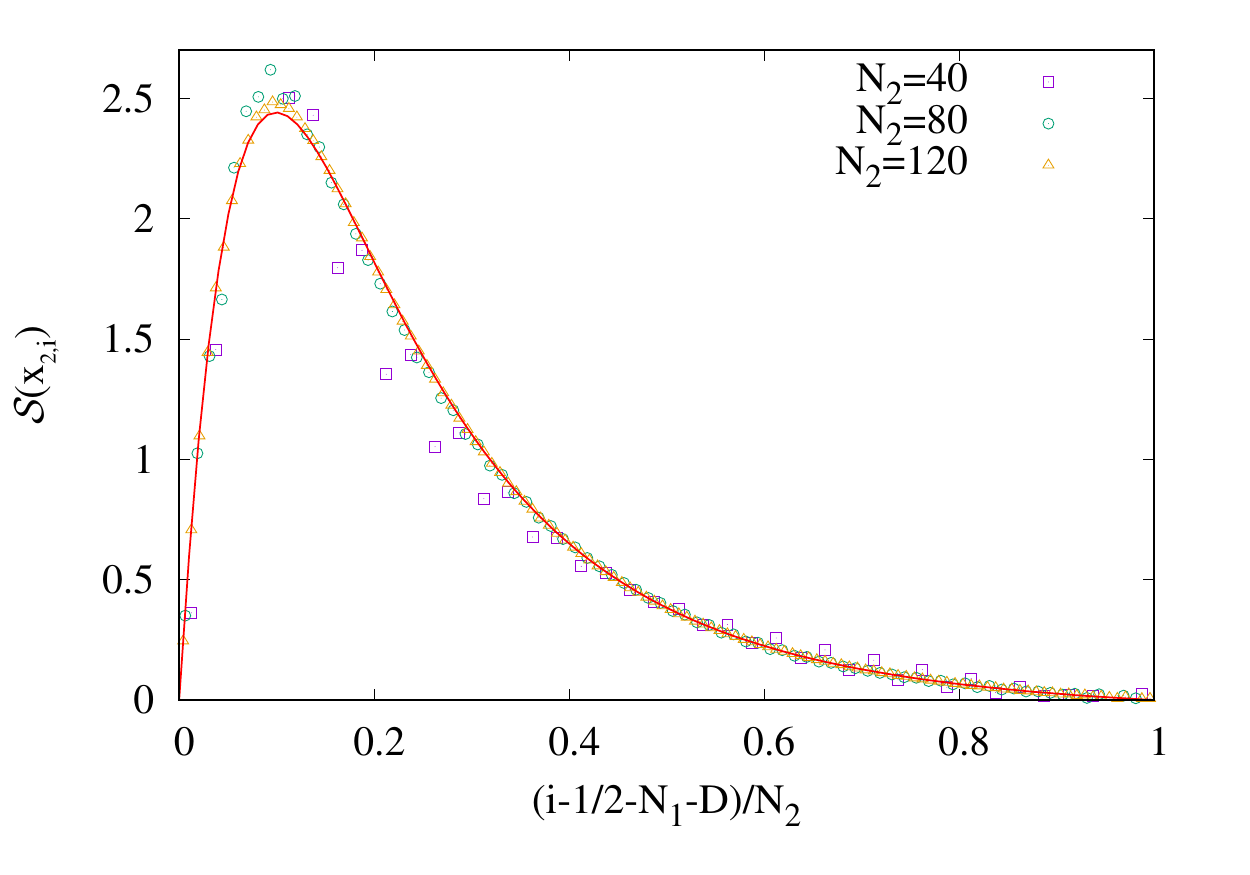}
\caption{
Continuum limit for unequal intervals. Top panel: $v(x_{\sigma,i})$ for the left and right intervals, respectively,
compared to $2\pi \beta(x)$. Bottom panel: $\mathcal{S}(x_{\sigma,i})$ for the left and right intervals, compared to $2\pi \betabix$.
The data is shown for various segment sizes, with the ratios $N_1/N_2=1/2$ and $D/N_2=1/8$ kept fixed.}
\label{fig:vs_unequal}
\end{figure}
%
%%%%%%%%%%%%%%%%%%%%%%%%%%%%%%%%%%%%%%%%%%%%%%%%%%%%%%%%%%%

Clearly, there are now two scales in the problem and by
taking the continuum limit one has to fix $N_\sigma s = \ell_\sigma$, with the boundary
coordinates given by $a_1=0$, $b_1=N_1 s$, $a_2=(N_1+D)s$ and $b_2=(N_1+N_2+D)s$.
Furthermore, the diagonal blocks also scale with the corresponding segment size
and the densities can be introduced as 
\eq{
h^{\textrm{\tiny $(\sigma)$}}_{i-p,i+p+1} \,=\,  -\,H^{\textrm{\tiny $(\sigma)$}}_{i-p,i+p+1}/N_\sigma \, .
}
However, the quantities $\beta(x)/\ell_\sigma$ and $\tilde \beta(x)$ are dimensionless and scale invariant,
i.e. they remain unchanged under a rescaling of each length in the functions. This property can be
used to fix either the length scale $\ell_1=1$ for the left interval or $\ell_2=1$ for the right one, leading to the definition
\eq{
v(x_{\sigma,i})= 2\sum_{p=0}^P (-1)^p (2p+1) \,h^{\textrm{\tiny $(\sigma)$}}_{i-p,i+p+1} \,,
\;\;\; \qquad \;\;\;
x_{\sigma,i} = i/N_\sigma \, .
\label{vxi2}  
}
The parameters of the corresponding $\beta(x)$ function must be scaled accordingly,
and can be shown to depend only on the two ratios $\nu =N_1/N_2$ and $r=D/N_2$.
In particular, for the right interval one has $a_1=0$, $b_1=\nu$, $a_2=\nu+r$ and $b_2=\nu+r+1$,
whereas the parameters for the left interval read $a_1=0$, $b_1=1$, $a_2=1+r/\nu$ and
$b_2=1+(1+r)/\nu$. The same arguments apply also for the bi-local weight,
which is defined analogously to \eqref{sxihf}, with arguments $x_{\sigma,i}=(i-1/2)/N_\sigma$.

The scaling functions of the local $v(x_{\sigma,i})$ and bi-local $\mathcal{S}(x_{\sigma,i})$ weights
are shown in Fig.\,\ref{fig:vs_unequal} for increasing segment sizes, keeping the ratios
$\nu=1/2$ and $r=1/8$ fixed. The results for the right interval are shifted such that they
also fall within the interval $(0,1)$. As in the previous examples, one can see a very good
convergence towards the expected continuum result. The only sizable deviations occur
for the shortest segments considered, corresponding to a distance of only $D=5$ lattice sites.
It is straightforward to generalize the calculation to arbitrary fillings, where one obtains again
a very good agreement with the same weight functions.

%\newpage
%%%%%%%%%%%%%%%%%%%%%%%%%%%%%%%%%%%%%%%%%%%%%%%%%%%%%%%%%%%%%%%%%%%%%%%%%%%%%%%%%%%%%%%%%%%%%%%
\section{Interval on the half line}
%%%%%%%%%%%%%%%%%%%%%%%%%%%%%%%%%%%%%%%%%%%%%%%%%%%%%%%%%%%%%%%%%%%%%%%%%%%%%%%%%%%%%%%%%%%%%%

Another important example where the entanglement Hamiltonian takes the form (\ref{H_A-def-sum-loc-bi-loc})
corresponds to the bipartition of the half line $x \geqslant 0$ where $A=(a,b)$ is an interval in a generic position. 

\subsection{Continuum results}
\label{sec_bdy_cft}

For the massless Dirac field, the most general boundary condition ensuring the global energy conservation
corresponds to imposing the vanishing of the energy flow, $T_{10}(x=0)=0$, through the boundary \cite{Cardy84, Cardy86, Cardy89}.
This boundary condition can be satisfied in two inequivalent ways, corresponding to a vector and
an axial $U(1)$ symmetry \cite{Liguori/Mintchev98}. Here we discuss only the former case, which corresponds to the
following boundary condition on the fields
\be
\label{vector-bc}
\psi_{\textrm{\tiny R}} (0) = \textrm{e}^{\textrm{i} \alpha}\, \psi_{\textrm{\tiny L}} (0)  \,,
\ee
with a scattering phase $\alpha \in [0,2\pi)$. Thus, the boundary condition at $x=0$ 
provides a scale invariant coupling of the fields having different chirality. 

The entanglement Hamiltonian of an interval $A=(a,b)$ on the half line $x \geqslant 0$
has been found in \cite{Mintchev/Tonni21} and it takes the form (\ref{H_A-def-sum-loc-bi-loc}).
In particular, the local term is still given by \eqref{general} with the weight function (\ref{beta-loc-2int-sym})
and the energy density $T_{00}$ defined in (\ref{T00-def}).
Instead, the bi-local term in (\ref{H_A-def-sum-loc-bi-loc}) is
\be
\label{K_A-bilocal-def-bdy}
\mathcal{H}_{\textrm{\tiny bi-loc}} 
\,=\,
2\pi 
\int_a^b \textrm{d} x  \,
\tilde{\beta}(x) \, T_{\textrm{\tiny bi-loc}}(x, \tilde{x}  ;\alpha) \, ,
\ee
where the weight function coincides with (\ref{beta-bi-loc-2int-sym}),
but the bi-local operator is defined as 
\be
\label{T-bilocal-def}
T_{\textrm{\tiny bi-loc}}(x, \tilde x ;\alpha) 
\equiv
\frac{\textrm{i}}{2}\;
\!:\!\!\Big[\,
\textrm{e}^{\textrm{i} \alpha} \, 
\psi^\dagger_{\textrm{\tiny R}}(x) \,  \psi_{\textrm{\tiny L}}(\tilde x)
- \textrm{e}^{-\textrm{i} \alpha} \,
\psi^\dagger_{\textrm{\tiny L}}(x) \,  \psi_{\textrm{\tiny R}}(\tilde x) +\textrm{h.c.} \,\Big] \!\!:  \;,
\ee
which depends on the phase occurring in the boundary condition \eqref{vector-bc},
and $\tilde{x}=ab/x$ is the point conjugate to $x$ inside $A$.
It is related to the conjugate point (\ref{x-conjugate-2-int}) for two disjoint intervals in the
symmetric configuration where $(a_1, b_1)=(-b,-a)$ and $(a_2, b_2)=(a,b)$.

\subsection{Lattice results}
\label{sec_bdy_lattice}

Consider now the lattice model given by a semi-infinite fermionic hopping chain, 
whose Hamiltonian is (\ref{Hff}) with the sums over $n$ restricted to $n \geqslant 1$,
and the subsystem $A=\left[d+1,d+N\right]$ made by $N$ consecutive sites and separated
by $d$ sites from the boundary of the chain.
The open half-chain is the simplest realization of the Dirac theory with a boundary,
corresponding to the choice $\alpha=\pi$. As suggested by the continuum
results above, the entanglement Hamiltonian should be very closely related to
the case of the symmetric double interval. We first derive the exact correspondence
on the lattice level.

%In order to investigate  the entanglement Hamiltonian of $A$,
Let us consider the symmetric double interval on the infinite chain,
composed of $A_2=\left[d+1,d+N\right]$ and its reflection to negative sites
 $A_1=\left[- \,d - N , -\,d - 1\right]$,  as in Sec.\,\ref{sec:equalint}.
The eigenvalues $\zeta_k$ and eigenvectors $\phi_k$ of the $2N\times 2N$
reduced correlation matrix with elements \eqref{corr-lattice} then follow from the equations
\be
\label{eigen-eq-def}
%\Bigg\{
\begin{array}{l}
C_{i',\,j'} \, \phi^{\textrm{\tiny (1)}}_k(j') + C_{i',j} \, \phi^{\textrm{\tiny (2)}}_k(j)  \,=\, \zeta_k\, \phi^{\textrm{\tiny (1)}}_{k}(i') \,,
\\ 
C_{i,\,j'} \, \phi^{\textrm{\tiny (1)}}_k(j') + C_{i,j} \, \phi^{\textrm{\tiny (2)}}_{k}(j)  \,=\, \zeta_k\, \phi^{\textrm{\tiny (2)}}_{k}(i) \,,
\end{array}
\ee
where $i,j \in A_2$ and $i',j' \in A_1$ and the $k$-th eigenvector has been decomposed into
two vectors corresponding to amplitudes on the left/right interval, respectively:
\eq{
\phi_k = 
\Bigg(
\begin{array}{c}
\phi^{\textrm{\tiny (1)}}_k \\
\phi^{\textrm{\tiny (2)}}_k
\end{array}
\Bigg) \,.
}
Due to the reflection symmetry of the geometry, the eigenvectors are either even or odd,
$\phi^{\textrm{\tiny (1)}}_k(-i)=\pm \phi^{\textrm{\tiny (2)}}_k(i)$. Using this property and
introducing $j'=-j$, $i'=-i$, the eigenvalue equations \eqref{eigen-eq-def} can be
decomposed into two sets
\be
C^{\pm}_{i,j} \phi^{\textrm{\tiny (2)}}_{k}(j)  \,=\, \zeta^\pm_k\, \phi^{\textrm{\tiny (2)}}_{k}(j) \, ,
\label{Cpm}
\ee
where the $N \times N$ correlation matrices $C^{\,\pm}$ with elements $i,j \in A_2$ are defined as
\be
\label{C-pm-mat-elements}
C^{\pm}_{i,j} = \frac{\sin [q_{\textrm{\tiny F}}(i-j)]}{\pi (i-j)} \pm 
\frac{\sin [q_{\textrm{\tiny F}}(i+j)]}{\pi (i+j)} \, .
\ee
In fact, the piece containing $i+j$ originates from the terms $C_{i,-j}$ and $C_{-i,j}$ in  \eqref{eigen-eq-def}
after making the above substitutions, whereas $C_{-i,\,-j}  = C_{i,j}$.
Note that the above observations are independent of the distance $d$.

Remarkably, the correlation matrices $C^{\pm}$ in \eqref{C-pm-mat-elements} are nothing
but the ones corresponding to a half-chain with Dirichlet or Neumann boundary conditions.
They can be obtained by replacing the plane-wave basis of the infinite chain with
appropriate standing-wave bases as
\be
C^{+}_{i,j} =
\frac{2}{\pi}
\int_0^{q_{\textrm{\tiny F}}} \! \cos(q\,i) \, \cos(q\,j)\, \textrm{d}q \, ,
\;\; \qquad \;\;
C^{-}_{i,j} =
\frac{2}{\pi}
\int_0^{q_{\textrm{\tiny F}}} \! \sin(q\,i) \, \sin(q\,j)\, \textrm{d}q \, .
\ee
Hence, the eigenvalue equation \eqref{Cpm} simply tells us, that the even/odd
eigenvectors and corresponding eigenvalues of the double interval problem are, up to normalization,
\emph{identical} to the ones for the half-chain with Dirichlet/Neumann boundary conditions.

%%%%%%%%%%%%%%%%%%%%%%%%%%%%%%%%%%%%%%%%%%%%%%%%
\begin{figure}[t!]
\vspace{-.3cm}
\centering
\includegraphics[width=0.45\textwidth]{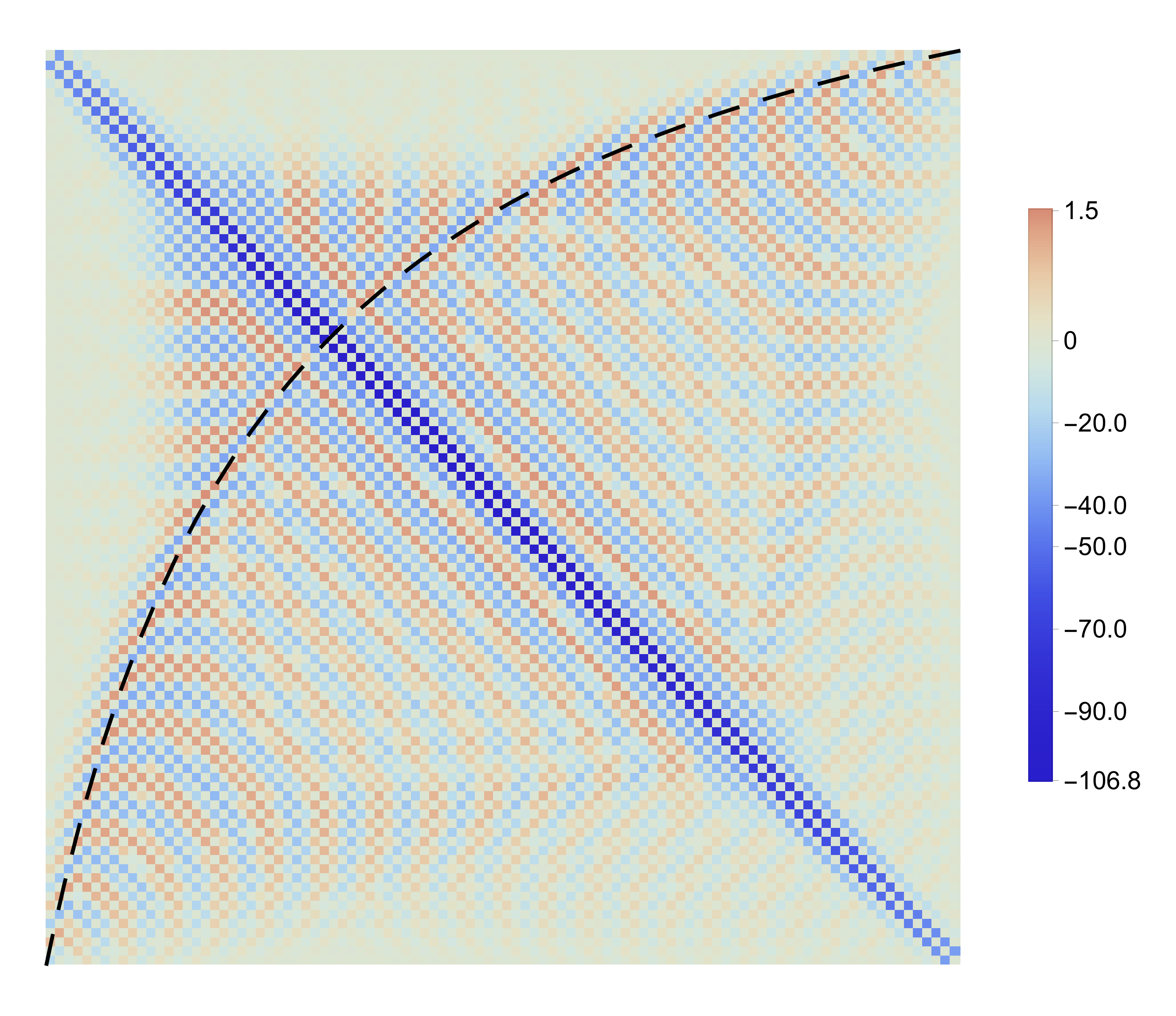}
\qquad
\includegraphics[width=0.45\textwidth]{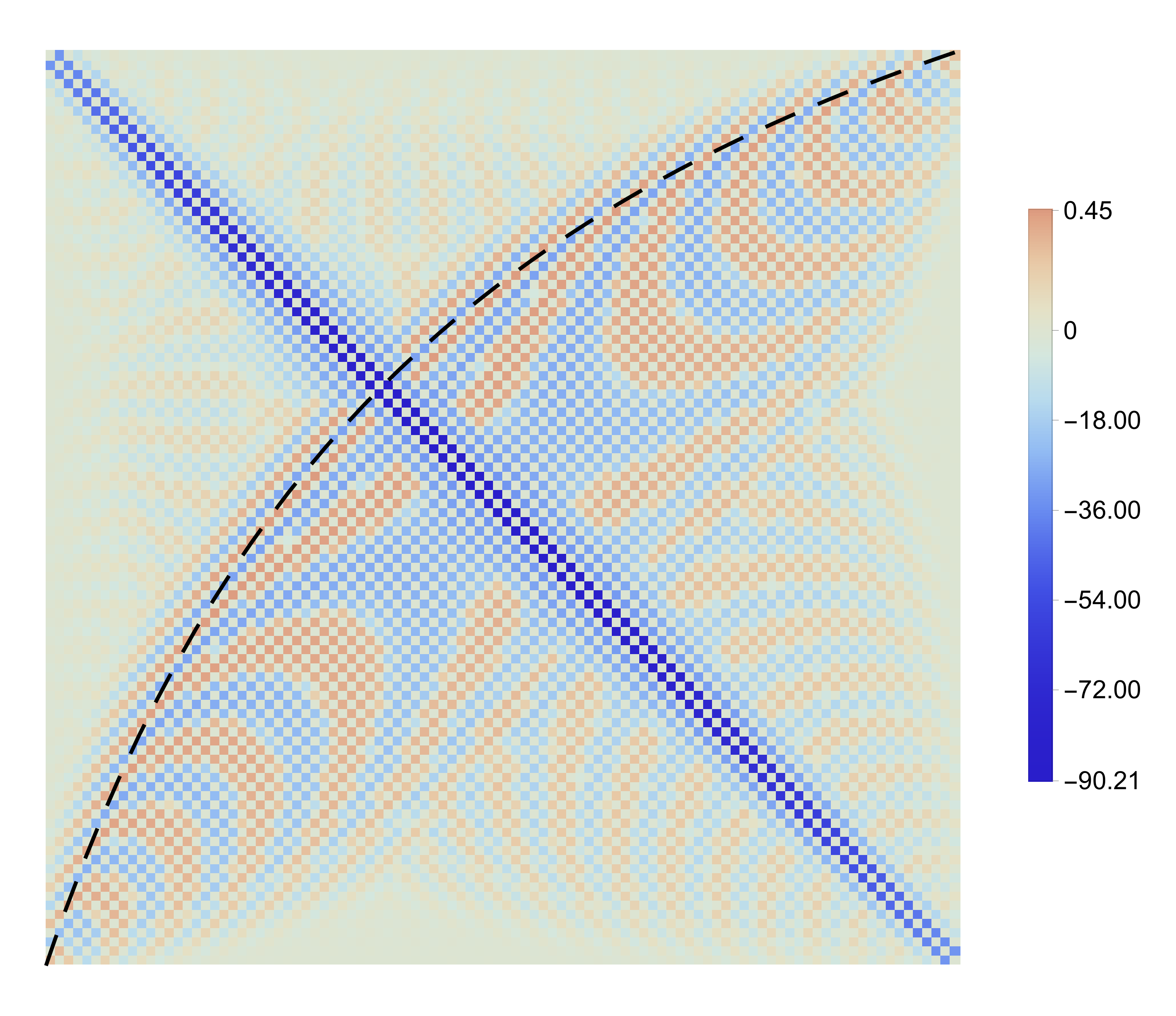}
\caption{
Entanglement Hamiltonian matrix for a block of consecutive sites 
in the semi-infinite chain with Dirichlet boundary conditions and at half filling. 
Left: $(d,N) = (25, 100)$. Right: $(d, N) = (50, 100)$.
The panels should be compared to the middle 
and bottom panel on the left of Fig.\,\ref{fig:EHmatrix}.
}
\label{fig:EHmatrixBdy1int-matrixplot}
\end{figure}
%%%%%%%%%%%%%%%%%%%%%%%%%%%%%%%%%%%%%%%%%%%%%%%%

The above observations can be used directly in the entanglement Hamiltonians
$H^\pm$ of the chains with corresponding boundary conditions. These are defined
analogously to \eqref{Hij} as
\eq{
H^{\pm}_{i,j}= 2 \sum_{k=1}^{N}\,\phi^{\pm}_k(i)\; \varepsilon^{\pm}_k\; \phi^{\pm}_k(j) \, ,
\;\;\;\; \qquad \;\;\;\;
i,j \in A_2\,,
\label{Hpm}}
such that they are composed of only the even/odd eigenvectors $\phi^{\pm}_k$ of the double interval problem,
with the factor two resulting from the different normalization. Comparing now to the block matrix
\eqref{Hblock}, one can see immediately that $H^+_{i,j} + H^-_{i,j} = 2H^{\textrm{\tiny (2)}}_{i,j}$,
as well as $H^+_{i,j} - H^-_{i,j} = 2H^{\textrm{\tiny (2,1)}}_{i,-j}$. In other words,
the entanglement Hamiltonians for the half-chain with Dirichlet/Neumann boundary conditions
can be written as
\be
H^{\pm}_{i,j}
=
H^{\textrm{\tiny (2)}}_{i,j} \pm {H}^{\textrm{\tiny (2,1)}}_{i,-j}\; ,
%H^{-}_{i,j}
%=
%H^{\textrm{\tiny (2)}}_{i,j} - {H}^{\textrm{\tiny (2,1)}}_{i,-j}
\label{HpmH2int}
\ee
in terms of the diagonal and off-diagonal blocks in the symmetric double interval.

In Fig.\,\ref{fig:EHmatrixBdy1int-matrixplot} we show $H^-$ 
corresponding to a block of $N=100$ sites and for two different values of the distance $d$
from the boundary of the open semi-infinite chain, in the special case of half filling.
These plots should be compared to the middle and bottom left panels of
Fig.\,\ref{fig:EHmatrix}, displaying the symmetric double interval with the corresponding $d$.
One can immediately recognize that the matrix $H^-$ for the open half-chain is indeed a kind
of superposition of the diagonal and off-diagonal blocks for the double interval,
as dictated by \eqref{HpmH2int}. In particular, one can see that the hyperbola
is now reflected and appears at $\tilde x=ab/x$, as shown by the black dashed lines in
Fig.\,\ref{fig:EHmatrixBdy1int-matrixplot}.

The result \eqref{HpmH2int} already makes it clear that the relation between the
half-chain problem and the double interval is completely analogous to the continuum
case in sec. \ref{sec_bdy_cft}. However, it remains to understand the difference
between the bi-local operators \eqref{T-bilocal-def} and \eqref{T-bilocal-2int},
by considering the continuum limit of $H^\pm$.
We first apply the substitutions \eqref{cpsi}, where the left/right-moving fields are
coupled via the boundary condition \eqref{vector-bc} at $x=0$. This coupling can
be made explicit by writing the mode expansion of the fields
$\psi_{\textrm{\tiny L}}(x)$ and $\psi_{\textrm{\tiny R}}(x)$,
where the parameter $\alpha$ enters as a relative phase between the left/right
propagating Fourier components. In our case here $H^+$ and $H^-$
correspond to the choice $\alpha=0$ and $\alpha=\pi$, respectively.
The change of the bi-local operator is due to the fact that the corresponding
matrix component in \eqref{HpmH2int} is mirrored. As pointed out already below
\eqref{bilochop}, it is not a priori clear which phase factors give a proper continuum 
limit in this expansion. Due to the appearance of ${H}^{\textrm{\tiny (2,1)}}_{i,-j}$,
it is now clear that for the half-chain one has to choose the terms with $i+j$.
This leads to the following expression
\eq{
\mathcal{H}^{\pm}_{\textrm{\tiny bi-loc}} =
\pm \int_{A} \dd x \,
\Big[ \,\mathcal{C}(x) \,  T_{\textrm{\tiny bi-loc}}^{\textrm{\tiny $(1)$}}(x,\tilde x)+
\mathcal{S}(x) \, T_{\textrm{\tiny bi-loc}}^{\textrm{\tiny $(2)$}}(x,\tilde x)\,\Big] \, ,
\label{EH12cont-pm}}
where the operators are defined as
\bea
T_{\textrm{\tiny bi-loc}}^{\textrm{\tiny $(1)$}}(x,\tilde x) 
&=& 
\frac{1}{2} \left[\,
\psi^\dagger_\textrm{\tiny R}(x) \, \psi_\textrm{\tiny L}(\tilde x) +
\psi^\dagger_\textrm{\tiny L}(x) \, \psi_\textrm{\tiny R}(\tilde x)  +\textrm{h.c.} 
\,\right] ,
\\
\rule{0pt}{.7cm}
T_{\textrm{\tiny bi-loc}}^{\textrm{\tiny $(2)$}}(x,\tilde x) 
&=&
 \frac{\mathrm{i}}{2} \left[\,
\psi^\dagger_\textrm{\tiny R}(x) \, \psi_\textrm{\tiny L}(\tilde x) -
\psi^\dagger_\textrm{\tiny L}(x) \, \psi_\textrm{\tiny R}(\tilde x)  +\textrm{h.c.} 
\,\right] ,
\eea
and the bi-local weights are given by
\eq{
\mathcal{C}(x_i) = \sum_{j \in A} \cos [q_Fs(j+i)] \, H^{\textrm{\tiny (2,1)}}_{i,-j} \; , 
\;\;\; \qquad \;\;\;
\mathcal{S}(x_i) = -\sum_{j \in A} \sin [q_Fs(j+i)] \, H^{\textrm{\tiny (2,1)}}_{i,-j} \; .
\label{scxp}}
Interchanging $j \to -j$, one clearly arrives at the same definition of the weights as in
\eqref{scx} for the double interval. Hence, in the limit $N \to \infty$, one has $\mathcal{S}(x_i) \to 2\pi \tilde \beta(x)$
as well as $\mathcal{C}(x_i) \to 0$, perfectly reproducing the result \eqref{K_A-bilocal-def-bdy}-\eqref{T-bilocal-def}
for the boundary Dirac theory with $\alpha=0,\pi$.

The case $0<\alpha<\pi$ can also be studied on the lattice by introducing a chemical potential
at the first site, i.e. adding the term $\mu_1 c_1^\dag c_1$ to the Hamiltonian with $\mu_1>0$.
The eigenvectors are then cosine functions with a phase shift
\eq{
\Phi(n) = \sqrt{\frac{2}{\pi}} \cos [q n + \delta(q)] \,, 
\;\;\; \qquad \;\;\;
\tan \delta(q) = \frac{1+2\mu_1 \cos q}{2\mu_1 \sin q}\,.
}
The parameter of the continuum theory then corresponds to the relative phase at the
Fermi level, $\alpha = 2\delta(q_F)$. It is easy to see that the correlation matrix has the form
\eq{
C^{\alpha}_{i,j} = \frac{\sin [q_{\textrm{\tiny F}}(i-j)]}{\pi (i-j)} +
\int_{0}^{q_F} \frac{\dd q}{\pi} \cos [q(i+j)+2\delta(q) ]\, ,
\label{Calpha}}
and the corresponding $H^\alpha$ can be evaluated numerically.
%Obviously, the drawback of the definition \eqref{scxp} is that it contains the bi-local matrix only.
However, in contrast to \eqref{HpmH2int}, for generic $\alpha$ it is not clear how to separate the local
and bi-local contributions that are superimposed in the matrix $H^\alpha$.

%%%%%%%%%%%%%%%%%%%%%%%%%%%%%%%%%%%%%%%%%%%%%%%%%%%%%%%%%%%
%
\begin{figure}[t!]
\center
%\hspace{-1.0cm}
\includegraphics[width=0.49\textwidth]{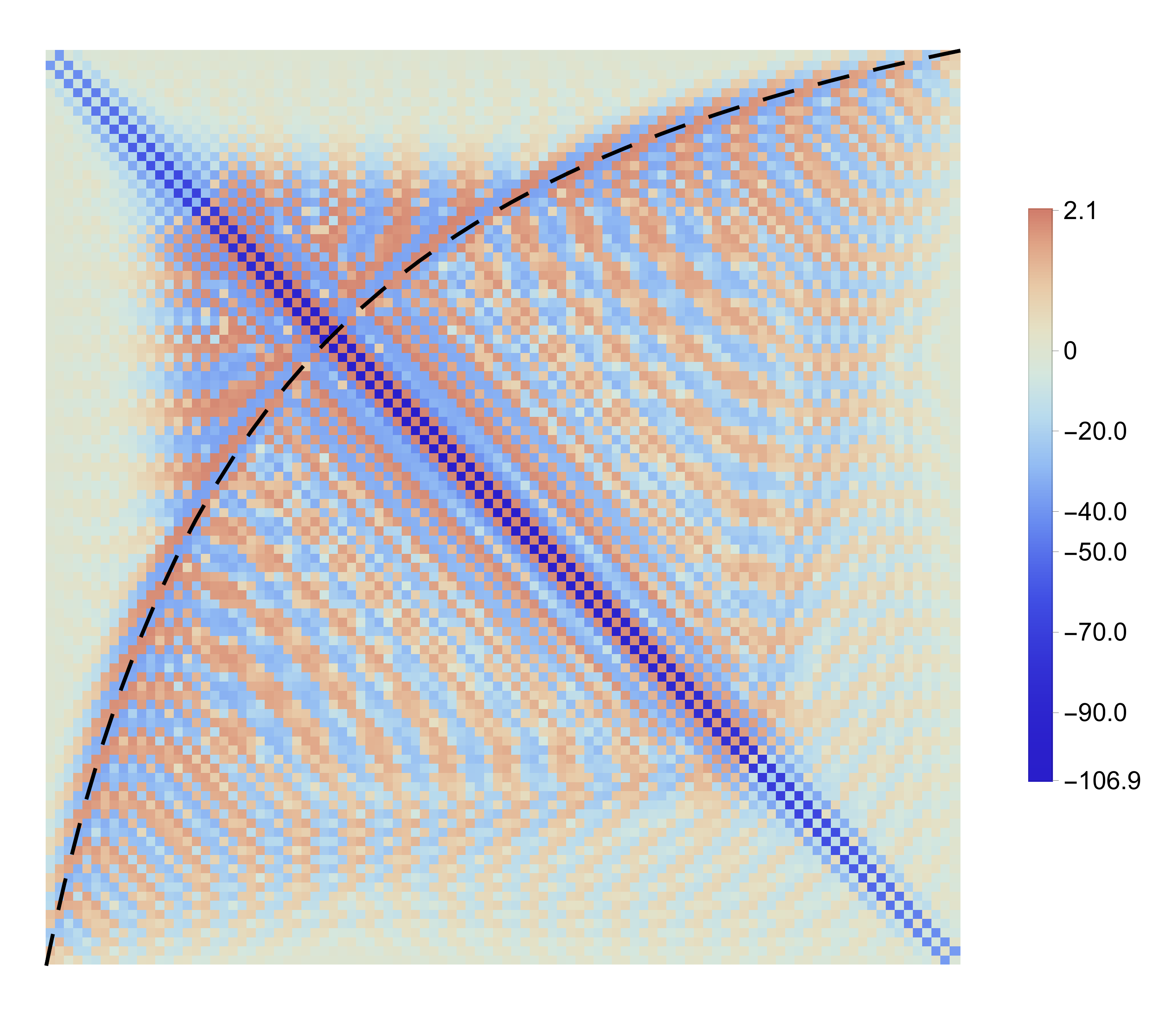}
\hspace{-0.5cm}
\includegraphics[width=0.52\textwidth]{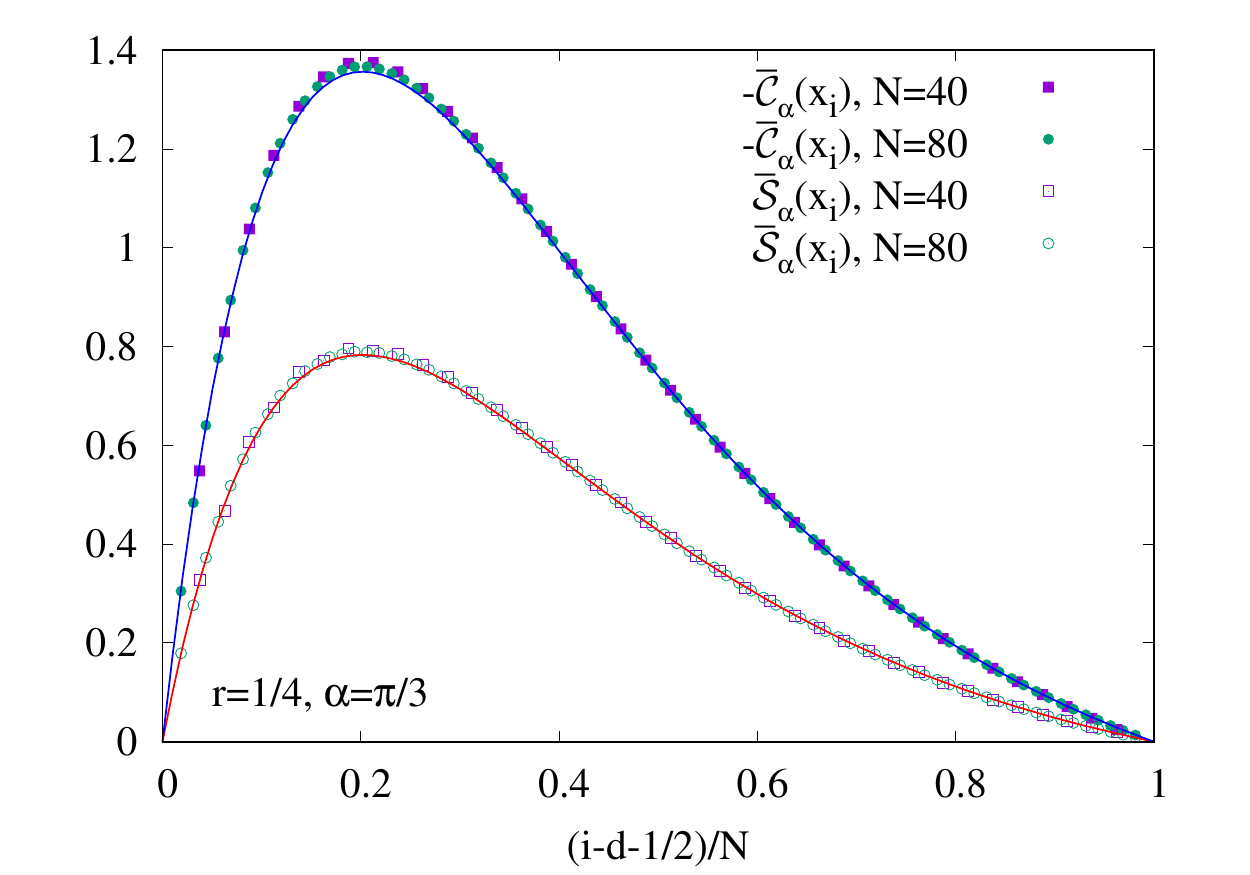}
\caption{Left: matrix plot for $d=25$, $N=100$ and $\alpha=\pi/3$. Right: bi-local weights from
\eqref{savg} for ratio $r=d/N=1/4$ and $\alpha=\pi/3$. The red and blue lines show the expected
continuum limit results in \eqref{barsc}.}
\label{fig:biloc_alpha}
\end{figure}
%
%%%%%%%%%%%%%%%%%%%%%%%%%%%%%%%%%%%%%%%%%%%%%%%%%%%%%%%%%%%

Here we propose such a method for the half-filled case. Inserting the full matrix
into the definition of the bi-local weights,
\eq{
\mathcal{C}_\alpha(x_i) = \sum_{j \in A} \cos\!\big(\tfrac{\pi}{2}(j+i)\big) \, H^\alpha_{i,j} \, ,
\;\;\; \qquad \;\;\;
\mathcal{S}_\alpha(x_i) = -\sum_{j \in A} \sin\!\big(\tfrac{\pi}{2}(j+i)\big) \, H^\alpha_{i,j} \, ,
\label{sct}}
the local contribution in $H^\alpha_{i,j}$ does not yield a proper
continuum limit, since the sum contains the phase factors with $i+j$. Indeed, one can argue
that this term has an alternating form $(-1)^i f(x_i)$ with some smooth function $f(x_i)$.
In order to get rid of this unwanted contribution, we define
\eq{
\bar{\mathcal{S}}_\alpha(x_i) = \mathcal{S}_\alpha(x_{i})/2 + \big[\mathcal{S}_\alpha(x_{i-1}) + \mathcal{S}_\alpha(x_{i+1})\big]/4\;,
\label{savg}}
and similarly for $\bar{\mathcal{C}}_\alpha(x_i)$. In other words, we combine the terms $i$ and $i\pm1$,
such that the alternating piece cancels up to second order corrections in the lattice spacing.
In order to reproduce the continuum result, one expects 
\eq{
\bar{\mathcal{C}}_\alpha(x_i) \,\to\, -\,2\pi \sin(\alpha) \, \tilde \beta(x)\,,
\;\;\; \qquad \;\;\;
\bar{\mathcal{S}}_\alpha(x_i) \,\to\, 2\pi \cos(\alpha) \, \tilde \beta(x) \,,
\label{barsc}}
to hold. For a generic scattering phase $\alpha$, the checkerboard structure of the matrix $C^\alpha$
in \eqref{Calpha} as well as of $H^\alpha$ is lost, and both of the sums in \eqref{sct} are nonzero.
This is demonstrated in Fig.\,\ref{fig:biloc_alpha}, where we show the results for $\alpha=\pi/3$
by choosing the value of $\mu_1$ accordingly. The matrix plot on the left is more blurred due
to the presence of nonzero elements $H^\alpha_{i,j}$ with even $j-i$, and on the right
one can see a very good convergence towards \eqref{barsc} for both bi-local weight factors.
We note that the local weights can be extracted analogously, inserting $H^{\alpha}$ into
\eqref{vmurow} and then defining the averages which now kill the bi-local contributions.
As expected, we find $\bar v_\alpha(x_i) \to 2\pi \beta(x_i)$ and $\bar\mu_\alpha(x_i) \to 0$ independently of $\alpha$.

%%%%%%%%%%%%%%%%%%%%%%%%%%%%%%%%%%%%%%%%%%%%%%%%%%%%%%%%%%%%%%%%%%%%%%%%%%%%%%%%%%%%%%%%%%%%%%%
\section{Commuting operators}
\label{sec_commuting}
%%%%%%%%%%%%%%%%%%%%%%%%%%%%%%%%%%%%%%%%%%%%%%%%%%%%%%%%%%%%%%%%%%%%%%%%%%%%%%%%%%%%%%%%%%%%%%

In the following we consider the integral operators whose kernels are the correlation functions
restricted to a subsystem $A$ and discuss differential operators that commute with them.

\subsection{Single interval}
\label{sec_commuting_single_interval}

We consider a single interval $A=(a,b)$ when the entire system is either on the line and in its
ground state, or on the circle and in its ground state, or on the line and in a thermal state.

The integral operator associated with the correlation function $G(x,y)$ restricted
to $A$ is
\be
\mathcal{I}_A[f](x) \equiv \int_A \textrm{d} y \; G(x,y)\, f(y)\, ,
\label{integral-op-def}
\ee
and depends both on $G(x,y)$ and on the spatial domain. 
Consider now the following first-order differential operator
%\cite{Casini/Huerta/09/disjoint, Mintchev/Tonni20-bdy}
%
\be
\mathcal{D}_A  \equiv  t(x)\, \frac{d}{dx} + \frac{1}{2}\,  t'(x) \, ,
\label{diff-op-def}
\ee
with a function $t(x)$ still to be determined and
evaluate the action of the operators (\ref{integral-op-def}) and (\ref{diff-op-def}) 
on a generic function $f$ in the two possible orders, namely
\be
\mathcal{D}_A \, \mathcal{I}_A[f] (x)
\,=
\int_A \textrm{d}y \big[ \mathcal{D}_{A,x}\,G(x,y) \big] f(y)
\,=
\int_A \textrm{d}y  \left[
t(x)\, \frac{d}{dx} G(x,y) +\frac{1}{2}\, t'(x)\, G(x,y)
\right] f(y)\, ,
\label{D-I-integral}
\ee
and
\be
\mathcal{I}_A \big[ \mathcal{D}_A f \big] (x)
\,=
\int_A\textrm{d}y  \left\{
- \,\frac{d}{dx} \big[ G(x,y)\, t(y) \big] + \frac{1}{2}\, G(x,y)\, t'(y)
\right\} f(y)
\,=\,
- \int_A  \textrm{d}y \big[ \mathcal{D}_{A,y}\,G(x,y) \big] f(y)\, ,
\label{I-D-integral}
\ee
with $\mathcal{D}_{A,x}$ denoting the operator (\ref{diff-op-def}), while
$ \mathcal{D}_{A,y}$ is (\ref{diff-op-def}) written in terms of the variabile $y$ instead
of $x$. In (\ref{I-D-integral}), an integration by parts has been performed with the
assumption that $t(x)$ vanishes at the endpoints of $A$, i.e.
\be
t(a)=t(b)=0\, .
\label{boundary-cond}
\ee
One then sees that the operators (\ref{integral-op-def}) and (\ref{diff-op-def}) commute
\be
\big[\,\mathcal{I}_A \, , \mathcal{D}_A \,\big] \,=\, 0\, ,
\label{commut-rel}
\ee
provided that
\be
\big( \mathcal{D}_{A,x}  + \mathcal{D}_{A,y} \big)G(x,y) = 0 \, .
\label{commut-cond}
\ee
Consider now the case of an infinite line with the system in its ground state. Then the kernel
is given by (\ref{corrfct-def})
\be
G(x,y) = \frac{\textrm{i}}{x-y} \, ,
\label{kernel-inf}
\ee
and the condition (\ref{commut-cond}) becomes explicitly
\be
\frac{t(x)-t(y)}{(x-y)^2} -\frac{t'(x)+t'(y)}{2(x-y)}=0\, .
\label{cond}
\ee
This functional equation is satisfied by an arbitrary quadratic function $t(x)$. The boundary condition
(\ref{boundary-cond}) then gives, up to a prefactor, 
\be
t(x)=(x-a)(b-x)\, .
\label{t-inf}
\ee
For this choice of $t(x)$, the differential operator (\ref{diff-op-def}) commutes with the integral
kernel and thus has the same eigenfunctions. We note that $t(x)$ is proportional to the quantity
$\beta(x)$ appearing in the entanglement Hamiltonian (\ref{general}).

With this result, one can easily obtain the solution for a system on a circle with circumference $L$
in its ground state. The kernel then is
\be
G(x,y) = \frac{\textrm{i}}{\sin [ \pi(x - y)/L]} \, ,
\label{kernel-circle}
\ee
and one can check that the generalization of (\ref{t-inf}) to sine functions
\be
t(x)=\sin[ \pi(x - a)/L]\,\sin[ \pi(b - x)/L] \, ,
\label{t-circle}
\ee
again satisfies (\ref{commut-cond}).
Finally, if the system is on a line but in a thermal state with inverse temperature $\beta$, one
has
\be
G(x,y) = \frac{\textrm{i}}{\sinh [ \pi(x - y)/\beta]} \, ,
\label{kernel-thermal}
\ee
and (\ref{commut-cond}) holds if one substitutes hyperbolic sine functions in (\ref{t-inf})
\be
t(x)=\sinh[(x - a)/\beta]\,\sinh[ (b - x)/\beta]\, .
\label{t-thermal}
\ee

The common eigenfunctions of $\mathcal{I}_A$ and $\mathcal{D}_A$ are easy to obtain from $\mathcal{D}_A$,
since this is a first-order differential operator. In all three cases one has
\be
\phi_s(x) \,=\, C\,\, \frac{\textrm{e}^{\textrm{i} s w(x)}}{\sqrt{t(x)}} \, ,
\label{phi-s}
\ee
with eigenvalue $\textrm{i} s$ and $w(x)$ given by
\be
w(x)=\int \frac{\textrm{d}x}{t(x)} \, .
\label{w-def}
\ee
For $t(x)$ given by (\ref{t-inf}), $w(x)$ is proportional to $\ln[(x-a)/(b-x)]$ and the exponential
factor describes the logarithmic oscillations found previously in \cite{Peschel04, Arias_etal17_1}.

\subsection{Modular flow}
\label{sec-mod-flow}

It was noted above that the function $t(x)$ in \eqref{t-inf} is proportional to the weight factor $\beta(x)$
in the entanglement Hamiltonian. In fact, the whole operator $\mathcal{D}_A$ of the previous subsection also
occurs in connection with $\mathcal{H}$, namely in the analysis of the modular flow of the chiral
components of the massless Dirac field. This is their unitary evolution via the (modular) Hamiltonian
$\mathcal{H}$ according to
\be
\label{mod-flow-psi}
\psi(x,\tau) \equiv \, \textrm{e}^{- \textrm{i}  \mathcal{H} \tau}\, \psi(x)\, \textrm{e}^{\textrm{i} \mathcal{H} \tau} \, ,
%\;\;\;\;\qquad\;\;\;\;
%\tau \in \mathbb{R}
\ee
where $\psi(x)$ corresponds to the field operator at $\tau=0$ and we dropped the index $R,L$ at
$\psi(x)$.

The equation of motion then is
\be
\textrm{i}\,\partial_\tau \psi(x,\tau) = \left[\,\mathcal{H},\psi(x,\tau) \right] \, ,
\label{equ-motion}
\ee
and inserting the expression for $\mathcal{H}$ given in \eqref{general} leads to the  
partial differential equation \cite{Casini/Huerta09, Mintchev/Tonni21}
\be
\label{mod-eq-psi}
\frac{1}{2\pi}\, \partial_\tau \psi(x,\tau) \,=\, \pm\,\mathcal{D}_{A} \,\psi(x,\tau) \, ,
\ee
where $\mathcal{D}_{A}$ is given by \eqref{diff-op-def} with $t(x) \equiv \beta(x)$
and $\pm$ corresponds to the right and left chirality, respectively.  
The expression on the right is the same as the one for finding the eigenfunctions $\phi(x)$ of the kernel of
$\mathcal{H}$ which have to be the same as those of the correlation kernel. Therefore, with the knowledge of 
$\mathcal{H}$, one could have found the commuting operator from \eqref{equ-motion}. However, in analogy to
other cases \cite{Peschel04,Eisler/Peschel13,Eisler/Peschel18,CNV19,CNV20}, where $\mathcal{H}$ is {\it{not}} known
in closed form, our aim in Sec.\,\ref{sec_commuting_single_interval} was to determine it directly from the
correlation function.

\subsection{Two disjoint intervals}
\label{sec_commuting_two_int}

It is possible to extend the considerations of
Sec.\,\ref{sec_commuting_single_interval} to the case where $A$ is the union of two disjoint intervals on
the line. When the entire system is in its ground state, the correlation
kernel is again given by (\ref{kernel-inf}) and one has to find an operator which takes into account
the non-local effects. This is somewhat difficult to guess, but one can invoke the equation of motion
\eqref{equ-motion}, since $\mathcal{H}$ is again known. In this way, one is lead to consider
\be
\label{diff-op-2int-def}
\tilde{\mathcal{D}}_A \,f(x) 
\equiv  
\left( t(x)\, \frac{d}{dx} + \frac{1}{2}\,   t'(x) \right) f(x) -  \tilde{t}(x) \, f(x_{\textrm{\tiny c}} ) \, ,
\ee
where the first piece is the operator $\mathcal{D}_A$ introduced in (\ref{diff-op-def}) and the second one
is an additional term where $f$ is evaluated at $x_{\textrm{\tiny c}} $, which is a function of $x$. 
For the moment, we leave the functions $t(x)$, $\tilde{t}(x)$ and $x_{\textrm{\tiny c}}(x)$ open, except for
the condition that $t(x)$ must now vanish at the endpoints of the two intervals. To obtain the commutator of
$\tilde{\mathcal{D}}_A$ and $\mathcal{I}_A$, one only has to consider the additional term, which gives
\be
\tilde{\mathcal{D}}_A \, \mathcal{I}_A[f] (x) \rightarrow 
\int_A \textrm{d}y 
 \left[\, - \tilde{t}(x)\, G(x_{\textrm{\tiny c}},y)
  \,\right] f(y)\, ,
\label{D-I-second}
\ee
and in reverse order
\be
\mathcal{I}_A \big[ \tilde{\mathcal{D}}_A f \big] (x) \rightarrow 
\int_A \textrm{d}y  \left[\,
  - \,G(x,y_{\textrm{\tiny c}})\, \tilde{t}(y_{\textrm{\tiny c}}) \, \frac{dy_{\textrm{\tiny c}}}{dy}
%  \partial_y y_{\textrm{\tiny c}}
\,\right] f(y)\, ,
\label{I-D-second}
\ee
where the integrand is the result of the change of integration variable $y \to y_{\textrm{\tiny c}}$
(with $y_{\textrm{\tiny c}}(y)$ defined analogously to $x_{\textrm{\tiny c}}(x)$) and
followed by a renaming of the integration variable.
If one now assumes that
\be
\tilde{t}(x_{\textrm{\tiny c}})\, \frac{dx_{\textrm{\tiny c}}}{dx}  \,=\, - \,\tilde{t}(x) \, ,
\label{prop-tilde-t}
\ee
this simplifies to
\be
\label{I-D-integral-2int-step1}
\mathcal{I}_A \big[ \tilde{\mathcal{D}}_A f \big] (x)
\rightarrow
\int_A  \textrm{d}y \left[\,
 \tilde{t}(y) \,G(x,y_{\textrm{\tiny c}})
\,\right] f(y)\, ,
\ee
which corresponds to (\ref{D-I-second}) but with the action on $y$ instead of $x$. Therefore, the
additional terms can be combined with the others and the condition for the vanishing of the commutator
is completely analogous to (\ref{commut-cond})
\be
\big( \tilde{\mathcal{D}}_{A,x}  + \tilde{\mathcal{D}}_{A,y} \big)G(x,y) = 0 \, .
\label{commut-cond-2}
\ee
It then remains to show that this relation holds for 
\be
t(x)= \beta(x)\,,\;\;\;\; \qquad  \;\;\;\; \tilde{t}(x)= \tilde{\beta}(x) \, ,
\label{choice}
\ee
where both quantities are given explicitly in (\ref{beta-loc-2int-sym}) and (\ref{beta-bi-loc-2int-sym})
for a symmetric double interval. Then $t(x)$ vanishes correctly at the boundaries $x=\pm a,\, \pm b$
and with $x_{\textrm{\tiny c}}=-ab/x$ one can verify easily that (\ref{prop-tilde-t}) holds.
The expression (\ref{commut-cond-2}) is more involved. The piece containing the function $t$ is
explicitly 
\be
\frac{t(x)-t(y)}{(x-y)^2} -\frac{t'(x)+t'(y)}{2(x-y)}\,=\,
\frac{(x^2-y^2)}{2(b-a)} \,\frac{ab(a+b)^2(ab-xy)}{(x^2+ab)^2(y^2+ab)^2} \, ,
\label{conj-1}
\ee
and does not vanish because $t(x)$ is not a quadratic function here, but it turns out that it
is compensated exactly by the contributions from $\tilde{t}$. Thus $\tilde{\mathcal{D}}_A$,
which might be called a functional differential operator, commutes with the integral operator
and can be used again to find the common eigenfunctions. This is sketched in Appendix\,\ref{sec_app_A}.

The commuting operator for the interval on the half line 
%(see Sec.\,\ref{sec_bdy_cft})
is discussed in Appendix\,\ref{sec_app_B}.

%\newpage
%%%%%%%%%%%%%%%%%%%%%%%%%%%%%%%%%%%%%%%%%%%%%%%%%%%%%%%%%%%%%%%%%%%%%%%%%%%%%%%%%%%%%%%%%%%%%
\section{Discussion}
\label{sec:conclusions}
%%%%%%%%%%%%%%%%%%%%%%%%%%%%%%%%%%%%%%%%%%%%%%%%%%%%%%%%%%%%%%%%%%%%%%%%%%%%%%%%%%%%%%%%%%%%%

We studied the entanglement Hamiltonian for two disjoint intervals in the hopping chain.
The diagonal part of the matrix, containing the hopping within each segment, has a similar
structure as for a single interval, with dominant nearest-neighbour terms. However,
their profile deviates increasingly from the expected weight function $\beta(x)$ 
as the distance between the intervals becomes smaller. In complete analogy to the single interval
case, the continuum result could be obtained by considering a proper continuum limit and including
the longer-range hopping into the definition of the local Fermi velocity. The hopping between the
segments is expected to have some relation to a bi-local term in the continuum model that couples
only to a single conjugate point. On the lattice, however, the landscape of the hopping amplitudes is
again more involved, with a clear ridge along the expected hyperbola but showing also some
extra features, see Fig.\,\ref{fig:EHmatrix}. Here, the continuum result for the bi-local weight was  
recovered by a proper row-wise sum of the matrix elements multiplied by an oscillatory factor.
Our numerical results show perfect agreement for arbitrary values of the filling
and of the segment sizes.

We also considered the problem of a single interval at some distance from the boundary
of a half-infinite chain. In the continuum, this setting is closely related to the case of two
symmetric intervals, and we find analogous relations for an open chain. Nontrivial boundary
conditions were implemented by a local chemical potential, and our continuum limit
perfectly reproduced the bi-local terms of the Dirac theory that depend on the scattering phase.

The continuum limit obtained for the bi-local terms could be used in a number of other situations.
In particular, one could apply it directly to the so-called negativity Hamiltonian 
\cite{Murciano/Vitale/Dalmonte/Calabrese22},
that is the analogous (albeit non-hermitian) object related to the partially time-reversed density
matrix for two disjoint intervals \cite{Shapourian/Shiozaki/Ryu17}. The continuum weight
functions then follow by applying a partial transposition as in \cite{Calabrese/Cardy/Tonni12}
to the double interval result.
One could also study bi-local terms in the entanglement Hamiltonian
of inhomogeneous chains, the simplest example being the case of a defect \cite{Mintchev/Tonni21-def}.
For slowly varying inhomogeneities, the continuum entanglement Hamiltonian for a single interval can be
obtained \cite{Tonni/Rodriguez-Laguna/Sierra17} via the curved-space CFT approach \cite{Dubail/Stephan/Viti/Calabrese17}. 
Recently it has been shown that the
continuum limit can be generalized to recover the local terms for the domain-wall melting problem \cite{Rottoli/Scopa/Calabrese22}.
It would be interesting to check whether this treatment could be extended to the bi-local terms
in case of two intervals. Further examples where non-local terms are expected
even for a single interval include the case of finite temperature and volume
\cite{Hollands19, Blanco/Perez-Nadal19, Fries/Reyes19},
systems with zero-modes \cite{Klich/Vaman/Wong15, Klich/Vaman/Wong15-long},
finite mass \cite{Arias_etal17_1,Eisler/DiGiulio/Tonni/Peschel20, Longo/Morsella20}
or systems driven out of equilibrium \cite{DiGiulio/Arias/Tonni19}.

One can also ask, how well the exact lattice entanglement Hamiltonian
can be approximated by a simpler one that does not contain all the long-range
hopping terms. For single intervals, one can simply replace the nearest-neighbour
hopping amplitudes by the properly discretized weights $\beta(x_i)$, which gives
a very accurate approximation of the exact reduced density matrix
\cite{Zhang/Calabrese/Dalmonte/Rajabpour20}.
The question then is how such an approximation would work for two intervals
and how to include the bi-local weights $\tilde \beta(x_i)$.
Studying these aspects, e.g. by variational methods as in \cite{Kokailetal21},
could shed some light on the role of these peculiar long-range couplings.

Somewhat related to this is the question of operators commuting with the correlation kernel (or
correlation matrix),
which we also considered. In the continuum, a simple first-order differential operator could be
given for a single interval, whereas for the double interval an additional long-range coupling appeared.
It would be quite interesting to know whether such a quantity also exists
in the lattice problem.

\section*{Acknowledgements}

\noindent
We would like to thank Mihail Mintchev and Pasquale Calabrese for discussions.
VE acknowledges funding from the Austrian Science Fund (FWF) through Project No. P35434-N.
ET's research has been conducted within the framework of the Trieste Institute for Theoretical Quantum Technologies (TQT).

%\newpage
%%%%%%%%%%%%%%%%%%%%%%%%%%%%%%%%%%%%%%%%%%%%%%%%%%%%%%%%%%%%%%%%%%%%%%%%%%%%%%%%%%%%%%%%%%%%%
\appendix
%%%%%%%%%%%%%%%%%%%%%%%%%%%%%%%%%%%%%%%%%%%%%%%%%%%%%%%%%%%%%%%%%%%%%%%%%%%%%%%%%%%%%%%%%%%%%

%%%%%%%%%%%%%%%%%%%%%%%%%%%%%%%%%%%%%%%%%%%%%%%%%%%%%%%%%%%%%%%%%%%%%%%%%%%%%%%%%%%%%%%%%%%%%
\section{Eigenfunctions of $\tilde{\mathcal{D}}_A$}
\label{sec_app_A}
%%%%%%%%%%%%%%%%%%%%%%%%%%%%%%%%%%%%%%%%%%%%%%%%%%%%%%%%%%%%%%%%%%%%%%%%%%%%%%%%%%%%%%%%%%%%%

The commuting operator $\tilde{\mathcal{D}}_A$ for the double interval is not a simple
object and leads to a functional differential equation for its eigenfunctions. We show here
that these can nevertheless be obtained with modest effort.

Using the single interval result $\phi_s(x)$ in \eqref{phi-s}, it is convenient to write them
in the form
\eq{
\phi_\pm(x) = m_\pm(x) \,\phi_s(x) = m_\pm(x) \,\frac{C}{\sqrt{t(x)}} \; \ex^{\mathrm{i} s w(x)} \,, 
\label{phipm}}
where we anticipated that they will turn out to be doubly degenerate, i.e. $\phi_\pm(x)$ 
both belong to the eigenvalue $\mathrm{i}s$.
%We assume the functions $m_\pm(x)$ to be real.
In order to ensure orthogonality, one has the condition
\eq{
\int_A \dd x \, \phi_+(x) \, \phi^*_-(x) = 0 \, .
\label{intphipm}}
With $t(x)>0$, this implies that $m_+(x)m_-(x)$ must change sign on the two intervals,
and we shall assume $m_+(x)>0$ as well as $m_-(x) \sim \mathrm{sign}(x-x_0)$.
Using the conjugate point, \eqref{intphipm} can be written as an integral on $A_1$ as
\eq{
\int_{A_1} \dd x \, \left[ \phi_+(x) \, \phi^*_-(x) + 
\phi_+(x_{\textrm{\tiny c}}) \,\phi^*_-(x_{\textrm{\tiny c}})\, \frac{d x_{\textrm{\tiny c}}}{d x} \right] = 0 \, .
}
This vanishes if
\eq{
\frac{\phi_+(x_{\textrm{\tiny c}})}{\phi_+(x)}\; \frac{\phi^*_-(x_{\textrm{\tiny c}})}{\phi^*_-(x)} 
= -\left(\frac{d x_{\textrm{\tiny c}}}{d x}\right)^{-1}
=-\frac{(x -x_0)^2}{R^2} \, .
}
Note that since $w(x)=w(x_{\textrm{\tiny c}})$, the exponential factors cancel (see (\ref{phipm})).
Owing to the degeneracy of the eigenfunctions, there should be some freedom in factorizing the above
condition. 
A simple consistent  choice turns out to be
\eq{
\frac{\phi_+(x_{\textrm{\tiny c}})}{\phi_+(x)} = 1 \, , 
\;\;\;\; \qquad \;\;\;\;
\frac{\phi^*_-(x_{\textrm{\tiny c}})}{\phi^*_-(x)} = -\frac{(x -x_0)^2}{R^2} \, .
\label{phiratio}}
Using the ansatz \eqref{phipm} in the differential equation, we obtain
\eq{
t(x) m'_{\pm}(x) \phi_s(x) = \tilde t(x) \phi_\pm(x_{\textrm{\tiny c}}) \, ,
}
which can be rewritten as
\eq{
\frac{m'_\pm(x)}{m_\pm(x)} = c(x)\, \frac{\phi_\pm(x_{\textrm{\tiny c}})}{\phi_\pm(x)} \, ,
\label{dmx}}
where we have introduced (for general disjoint intervals)
\eq{
c(x) = \frac{\tilde t(x)}{t(x)} = \frac{R^2}{(x-x_0)\left[(x-x_0)^2+R^2\right]} \, .
}
Let us first consider the eigenfunction $\phi_+(x)$ and substitute the corresponding ratio
\eqref{phiratio} into \eqref{dmx}. The equation can then be integrated as
\eq{
\ln \left[ m_+(x) \right] = \int \dd x \, c(x) = 
\frac{1}{2} \ln \left[ \frac{(x-x_0)^2}{(x-x_0)^2+R^2} \right],
}
such that one has
\eq{
  m_+(x) = \frac{|x-x_0|}{\sqrt{(x-x_0)^2+R^2}} \, .
\label{m-plus}
}
Using the property
\eq{
\frac{t(x_{\textrm{\tiny c}})}{t(x)}= -\frac{x_{\textrm{\tiny c}} -x_0}{x-x_0}= \frac{R^2}{(x -x_0)^2} \, ,
}
as well as $w(x_{\textrm{\tiny c}})=w(x)$, it is easy to verify that the ratio indeed gives one
\eq{
\frac{\phi_+(x_{\textrm{\tiny c}})}{\phi_+(x)} = 
\frac{m_+(x_{\textrm{\tiny c}})}{m_+(x)} \, \sqrt{\frac{t(x)}{t(x_{\textrm{\tiny c}})}} = 1 \, .
}
Analogously, for the other ratio in \eqref{phiratio} one can also
easily integrate \eqref{dmx} with the result
\eq{
  m_-(x) = \frac{\mathrm{sign}(x-x_0) \, R}{\sqrt{(x-x_0)^2+R^2}} \, ,
\label{m-minus}
}
where the sign function ensures that the ratio $\phi_-(x_{\textrm{\tiny c}})/\phi_-(x)$ is indeed negative.
The functions $m_{\pm}(x)$ are the same ones as found in \cite{Casini/Huerta09,Mush-book, Arias_etal18} after
orthogonalization.

The choice made in \eqref{phiratio} is the simplest one that leads to real-valued $m_\pm(x)$.
One can also see what happens if one assigns the ratios symmetrically
\eq{
\frac{\phi_+(x_{\textrm{\tiny c}})}{\phi_+(x)} = \,\textrm{i}\,\frac{x -x_0}{R} \, , 
\;\;\;\; \qquad \;\;\;\;
\frac{\phi^*_-(x_{\textrm{\tiny c}})}{\phi^*_-(x)} = \,\textrm{i}\,\frac{x -x_0}{R} \, .
}
Note that in order to have linearly independent solutions, we now need to work with
complex functions $m_\pm(x)$. The equation \eqref{dmx} can again be integrated and
for $x>x_0$ yields
\eq{
\ln [m_+(x)] =\, \textrm{i} \, \arctan\frac{x-x_0}{R } \, .
}
Using the identity 
\eq{
\textrm{i}\,\arctan z = \frac{1}{2} \ln \frac{1+\textrm{i}\, z}{1-\textrm{i} \,z}=\ln \frac{1+\textrm{i}\,z}{\sqrt{1+z^2}} \, ,
}
one arrives at
\eq{
m_+(x) = 
\frac{\textrm{i}\,|x-x_0|+\mathrm{sign}(x-x_0) R}{\sqrt{(x-x_0)^2+R^2}} \, .
}
This is clearly just a simple linear combination of the eigenfunctions found before.
It is easy to check that it satisfies $m_+(x_{\textrm{\tiny c}})/m_+(x)= \textrm{i} \, \mathrm{sign}(x-x_0)$ as it should,
and $m_-(x)=m^*_+(x)$.

%%%%%%%%%%%%%%%%%%%%%%%%%%%%%%%%%%%%%%%%%%%%%%%%%%%%%%%%%%%%%%%%%%%%%%%%%%%%%%%%%%%%%%%%%%%%%
\section{Commuting operator for the interval on the half line}
\label{sec_app_B}
%%%%%%%%%%%%%%%%%%%%%%%%%%%%%%%%%%%%%%%%%%%%%%%%%%%%%%%%%%%%%%%%%%%%%%%%%%%%%%%%%%%%%%%%%%%%%

Following the discussion of Sec.\,\ref{sec_commuting},
we construct here a commuting operator corresponding to the 
interval $A =(a,b)$ on the half line $x \geqslant 0$.

The non-vanishing two point functions for the free massless Dirac field on the half line
can be organised in a $2\times 2$ matrix, which depends on the boundary condition
(see \cite{Mintchev/Tonni21}). 
This correlation matrix leads us to consider the integral operator
\be
\label{integral-op-bdy-def}
\mathcal{I}_A[ \boldsymbol{f} ](x) \equiv \int_a^b   \textrm{d} y\;
\Bigg(\begin{array}{cc}
G_{-}(x,y)  & \textrm{e}^{\textrm{i} \alpha}\,G_{+}(x,y)
\\
- \,\textrm{e}^{-\textrm{i} \alpha}\,G_{+}(x,y)\;  & - \,G_{-}(x,y) 
\end{array}\Bigg)
\Bigg(\begin{array}{c}
f_1(y)  \\  f_2(y)
\end{array}\Bigg)\,,
\;\;\qquad\;\;
G_\mp(x,y) \equiv  \frac{\textrm{i}}{x \mp y}\,,
\ee
where $\alpha$ is the parameter entering in the boundary condition (\ref{vector-bc})
and $\boldsymbol{f}(x)$ denotes a vector whose components are two generic functions $f_1(x) $ and $f_2(x)$.
Furthermore, from the modular flow for this case studied in \cite{Mintchev/Tonni21}
we introduce the following vectorial functional differential operator
\be
\label{diff-op-bdy-def}
\mathcal{D}_A \,\boldsymbol{f}(x) 
\,\equiv  \,
\left(\begin{array}{c}
\displaystyle
%\mathscr{D}_x 
\left(  t(x)\, \frac{d}{dx} + \frac{1}{2}\,   t'(x) \right) f_1(x) 
-  \textrm{e}^{\textrm{i} \alpha} \,\tilde{t}(x) \, f_2(\tilde{x})
\\ 
\rule{0pt}{.8cm}
\displaystyle
 \textrm{e}^{- \textrm{i} \alpha} \,\tilde{t}(x) \, f_1(\tilde{x}) 
 - \left(  t(x)\, \frac{d}{dx} + \frac{1}{2}\,   t'(x) \right) f_2(x)
\end{array}
\right) ,
\ee
in terms of the functions $t(x) $, $\tilde{t}(x) $ and $\tilde{x}(x)$,  to determine under the condition $t(a) = t(b) =0$.

From a computation similar to the one discussed in Sec.\,\ref{sec_commuting_two_int} for two disjoint intervals on the line,
we find that  the operators (\ref{integral-op-bdy-def}) and (\ref{diff-op-bdy-def}) commute
provided that $\tilde{x}(x)=ab/x$, that
\be
\tilde{t}(\tilde{x})\, \frac{d\tilde{x}}{dx}  \,=\, - \,\tilde{t}(x)\,,
\label{conj-tilde-t}
\ee
and 
\be
%\label{dd-eq-bdy}
\left[\,
 t(x)\, \frac{d}{dx} + \frac{1}{2}\,   t'(x)  
 + 
 t(y)\, \frac{d}{dy} + \frac{1}{2}\,   t'(y) 
\, \right] G_{-}(x,y) 
+ \tilde{t}(x) \, G_{+}(\tilde{x},y)
 - \tilde{t}(y) \, G_{+}(x, \tilde{y}) 
=
0\,,
\ee
which  is equivalent to (\ref{commut-cond-2}) specialised to the case of two equal intervals on the line placed symmetrically 
with respect to the origin. 
Thus, the operators (\ref{integral-op-bdy-def}) and (\ref{diff-op-bdy-def}) commute 
when $t(x)= \beta(x)$ and $\tilde{t}(x) = \tilde{\beta}(x)$ are  the weight functions 
(\ref{beta-loc-2int-sym}) and (\ref{beta-bi-loc-2int-sym}), respectively.

\newpage
\section*{References}
%%%%%%%%%%%%%%%%%%%%%%%%%%%%%%%%%%%%%%%%%%%%%%%%%%%%%%%%%%%%%%%%%%%%%%%%%%%%%%%%%

\providecommand{\newblock}{}

\end{document}